\documentstyle[aps,prb,preprint]{revtex}  
\draft
\tighten
\begin{document}
\title{Electrically tunable GHz oscillations in doped \\
GaAs-AlAs superlattices}
\author{J. Kastrup\cite{JK}, R. Hey, and K.H. Ploog}
\address{Paul-Drude-Institut f\"ur Festk\"orperelektronik, \\
Hausvogteiplatz 5-7, D-10117 Berlin, Germany}

\author{H. T. Grahn\cite{HTG}}
\address{Research Center for Quantum Effect Electronics, Tokyo Institute
of Technology,\\
2-12-1 O-okayama, Meguro-ku, Tokyo 152, Japan}

\author{L. L. Bonilla, M. Kindelan, M. Moscoso, A. Wacker\cite{AW}}
\address{Escuela Polit\'ecnica Superior, Universidad Carlos III de
Madrid,\\
Butarque 15, E-28911 Legan\'es, Spain}

\author{J. Gal\'{a}n}
\address{Departamento de Matem\'atica Aplicada II,
Escuela Superior de Ingenier\'{\i}a,\\
Avda. Reina Mercedes s/n, 41012 Sevilla, Spain}


\maketitle

\begin{abstract}
Tunable oscillatory modes of electric-field
domains in doped semiconductor superlattices are reported.
The experimental investigations demonstrate the realization
of tunable, GHz frequencies in GaAs-AlAs superlattices covering
the temperature region from 5 to 300~K. The orgin of the
tunable oscillatory modes is determined using an
analytical and a numerical modeling of the dynamics of domain
formation. Three different oscillatory modes are found.
Their presence depends on the actual shape of the drift velocity
curve, the doping density, the boundary condition, and
the length of the superlattice. For most bias regions, the
self-sustained oscillations are due to the formation, motion,
and recycling of the domain boundary inside the superlattice.
For some biases, the strengths of the low and high field domain
change periodically in time with the domain boundary being pinned
within a few quantum wells. The dependency of the frequency on the
coupling leads to the prediction of a new type of tunable GHz
oscillator based on semiconductor superlattices.\\
\end{abstract}

\pacs{PACS numbers: 73.40.Gk, 73.50.Fq, 73.61.Ey}
\section{Introduction}

Since the observation and explanation of the Gunn-effect in the early 60's
it has been known that traveling field inhomogeneities are generated by
negative differential velocity (NDV).\cite{Ridley61,Gunn63} The properties
of these traveling Gunn-domains have been studied extensively theoretically
and experimentally \cite{Shur87,Higuera92} leading to several proposals for
applications. The most important application is probably the frequency
generator. Although oscillator devices based on the Gunn-effect operating in
the range of 100~GHz have been reported,\cite{Ruttan75} Gunn-oscillators
have not led to the expected breakthrough in technology. This is largely due
to the difficulties in tuning the oscillation frequency. Furthermore, it is
impossible to vary the drift velocity vs field characteristics ($v(F)$ curve)
except by using different bulk materials (e.g., GaAs, InGaAs or InP).

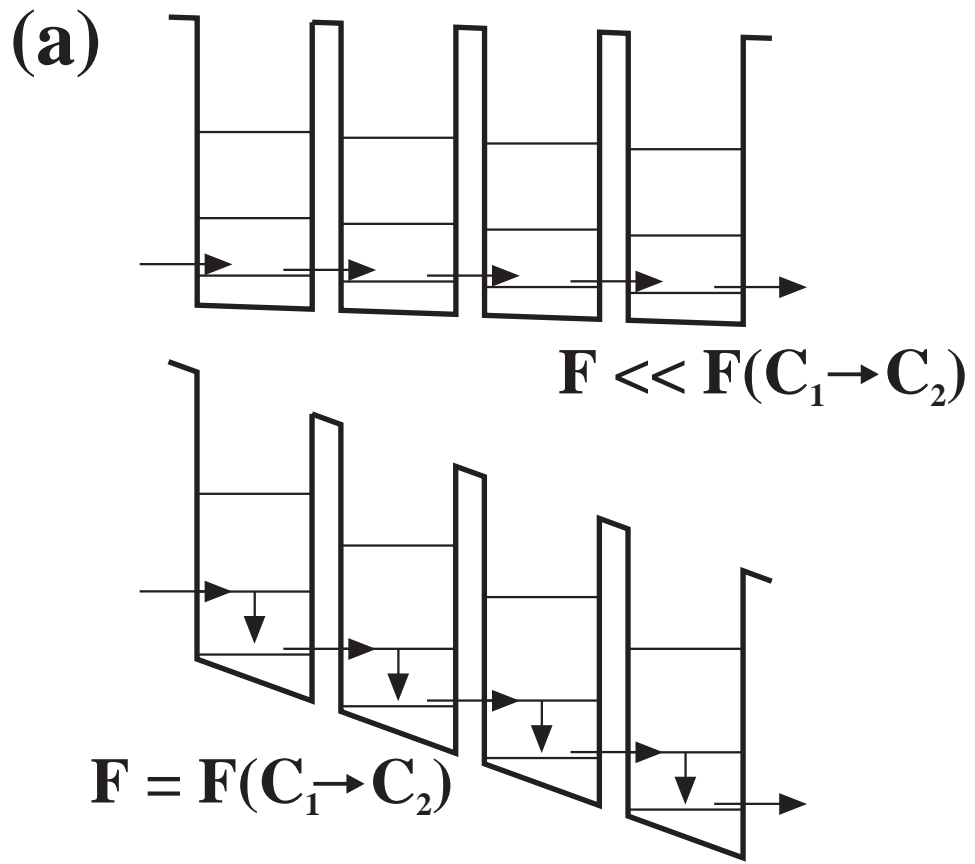
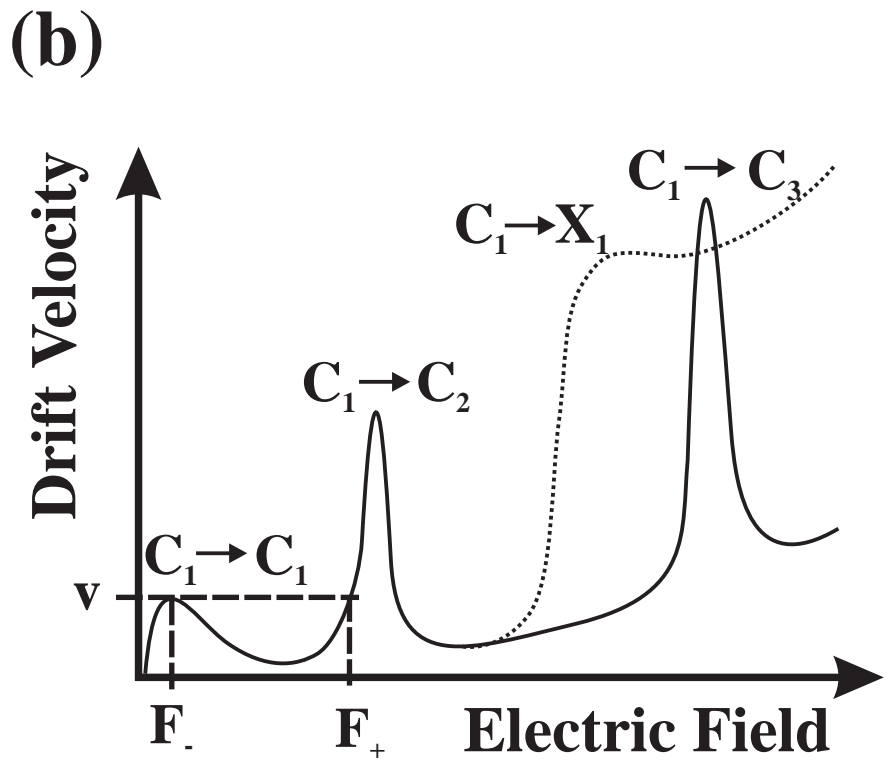
\begin{figure}
\caption{(a) Conduction band profile of a superlattice with three
subbands in an applied electric field. 
(b) Drift velocity vs field characteristics of this superlattice
consisting of the nonresonant background, the low-field maximum
($C_1 \rightarrow C_1$),
and the first two resonant tunneling maxima. The dotted line indicates a
possible shape of the $v(F)$ curve in the case of a $\Gamma \rightarrow X$ resonance
located between the $C_1 \rightarrow C_2$ and $C_1 \rightarrow C_3$ resonances.
The dashed line is explained in the text.}
\label{slwfield}
\end{figure}

These problems may be overcome by using an entirely different class of
NDV exhibiting materials formed by semiconductor superlattices. In an
applied electric field perpendicular to the two-dimensional layers, several
different transport mechanisms such as miniband transport, resonant
tunneling, and real-space transfer from well to barriers
($\Gamma \rightarrow X$ transport) give
rise to complex $v(F)$ curves with several regions of NDV as shown in
Fig.~\ref{slwfield}. In contrast to the
above mentioned materials, the tunability of the barrier and well thickness
as well as the control over the barrier height in the superlattice can be
used to tailor the drift velocity vs field $v(F)$ characteristics. By proper
engineering, samples with several NDV regions can be manufactured with
control over previously inaccessible features such as the actual shape of
the NDV (cf. Fig.~\ref{slwfield}). The frequency of the oscillations also
depends on the number of moving charges in the superlattice, which may be
controlled by varying the doping in the quantum wells.

Currently, three different types of superlattice oscillations due to NDV
have to be distinguished :

\begin{itemize}
\item  The most prominent type of oscillations is found, when carriers within
the miniband are accelerated beyond the Brillouin zone boundary, where their
drift velocity becomes negative. In the absence of scattering, the electron
wave packet then oscillates with the Bloch frequency $f_B=eFd/h$,
where $d$\ and $h$\ denote the superlattice period and
Planck's constant, respectively.\cite{Bloch28,Zener34}
These so-called Bloch oscillations were predicted for
superlattices by Esaki and Tsu in 1970.\cite{Esaki70} Inspired by the
prospect of THz frequency generators based on superlattice Bloch
oscillators, a long series of investigations targeting the miniband regime
followed leading to the observation of damped Bloch oscillations in
experiments with pulsed optical excitation a few years ago.\cite{Feldmann-PRB92}

\item  A different type of oscillations occurs in the miniband regime, when
the scattering times are shorter than the tunneling times. In this case
transient charge accumulations traveling through the superlattice may lead
to oscillations in the current. Experiments using pulsed optical excitation
of carriers in superlattices with wide minibands showed damped oscillations
with frequencies up to 20 GHz.\cite{LePerson92} More recently, reflection
gain up to 60~GHz in doped superlattices with miniband widths of up to
80~meV was demonstrated.\cite{Hadjazi93}

\item Very recently a third type of oscillations has been shown to exist in
weakly coupled superlattices in field regions where tunneling into
higher subbands takes place.\cite{Kastrup-PRB95,Grahn-jjap95,Kastrup-SSE95}
As a consequence of domain formation, this type of oscillations may also be
observed in field regions where the electrons tunnel between the lowest
subbands. While the first two oscillation types have never been observed
under constant bias conditions, this new type is shown to operate over a
broad frequency range practically independent of the external circuit and
without any external triggering.
\end{itemize}

In this paper we investigate experimentally and theoretically
the third type of oscillations in lightly and moderately doped superlattices
(SL). This type of oscillations may be generally observable
in the NDV regime of the $v(F)$ characteristics of such superlattices.
In a companion paper, \cite{WAC96} we have shown that self-sustained
oscillations may exist if the dimensionless doping parameter, $\nu = e N_{D}
d/(\epsilon F_{M})$ ($N_D$ is the 3D SL doping, and $F_M$ is the field at the
resonance we consider) lies between two values $(\nu_1/N,\nu_2)$ which can
be calculated numerically from the local $v(F)$ characteristics
($N$ is the number of SL periods). If
several tunneling resonances exist (cf. Fig.~\ref{slwfield}(a) and (b)) the
frequency of the oscillations increases with increasing index of the
subband involved in the tunneling resonance. This observation is attributed
to the larger drift velocities associated with the resonances involving
higher subbands. Furthermore, several new oscillatory modes without the
typically well-defined domains are shown to exist. These oscillatory
modes can be tuned in frequency simply by changing the applied voltage.
Because the oscillations are observable even at room temperature, we suggest
to use superlattice oscillators based on resonant tunneling into higher
subbands as a tunable source for high frequencies.

\section{Experimental Results}

Four GaAs/AlAs superlattice structures, grown by molecular beam epitaxy, are
discussed in this paper. The samples are hereafter referred to as 9.0/4.0
(doped and undoped), 9.0/1.5, and 13.3/2.7, where the first number refers
to the GaAs well width and the second to the AlAs barrier thickness in nm
(cf. Table~\ref{samples}). The superlattices are embedded between two highly
doped Al$_{0.5}$Ga$_{0.5}$As contact layers with a doping density
N$_D$=2$\times 10^{18}$~cm$^{-3}$ (Si for $n^{+}$- and Be for $p^{+}$-type
doping) forming $n^{+}$-$n$-$n^{+}$ and $p^{+}$-$i$-$n^{+}$diodes. After
evaporating AuGe/Ni contacts onto the top and substrate side and alloying
them for ohmic connections, the samples are wet-etched into mesas of 120~$\mu$m
diameter. The circular top contacts of 70~$\mu$m diameter leave a large part
of the mesas uncovered to allow for optical access. All experiments are
performed at 5~K in a He-flow cryostat using high-frequency coaxial cables
with a bandwidth of 20~GHz. The time-averaged current-voltage data are recorded
with a Keithley SMU236. The time-resolved current is detected with a Tektronix
sampling oscilloscope CSA~803 using the GHz sampling head SD-32.

\subsection{Current-voltage characteristics}

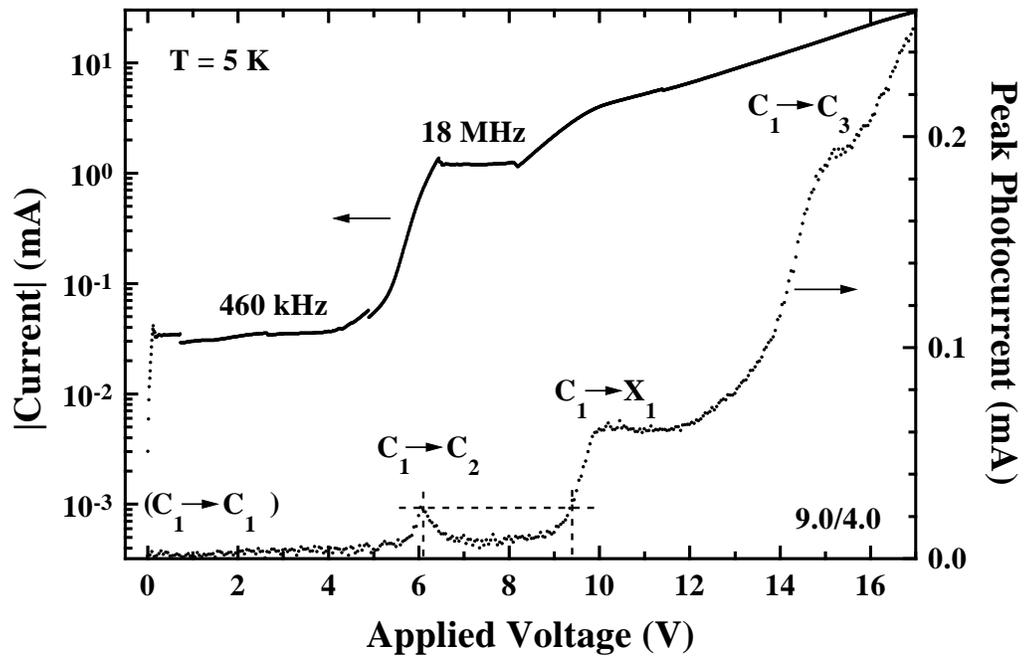
\begin{figure}
\caption{Current-voltage characteristics of the doped 9.0/4.0 sample and
peak photocurrent vs applied voltage obtained by time-of-flight experiments
in the undoped 9.0/4 reference sample.
The dashed line indicates the condition of current conservation for the
second plateau
assuming a low-field domain with a field strength corresponding to the
$C_1 \rightarrow C_2$ resonance.}
\label{IV6750}
\end{figure}

Fig.~\ref{IV6750} shows the current-voltage (I-V) characteristics of the
doped 9.0/4.0 sample in forward bias. Clearly visible are two plateau regions
with almost constant current between 0.5 and 4~\thinspace V and 6.5 and
8.5~\thinspace V. To prove that these plateaus are related to the subband
resonances, time-of-flight (TOF) measurements were performed on an
undoped 9.0/4.0 reference sample ($p^{+}$-$i$-$n^{+}$ diode).\cite{Schneider89}
The peak photocurrent is thereby taken as a measure for the inverse transport
time, which at low excitation densities is directly proportional to the
drift velocity for homogeneous fields. Note that for the presentation in
Fig.~\ref{IV6750} the built-in voltage ($1.5$~V) of the $p^{+}$-$i$-$n^{+}$
diode has been subtracted from the reverse bias voltage applied to the
$p^{+}$-$i$-$n^{+}$ diode and the sign has been inverted. The resonances
for tunneling from the first into the second ($C_1 \rightarrow C_2$) as
well as into the third ($C_1 \rightarrow C_3$) conduction subband are clearly
observed at 6 and 15~V, respectively. The low-field transmission maximum
($C_1 \rightarrow C_1$), which in the case of a strongly coupled superlattice
corresponds to miniband conduction, is not resolved in this sample, because the
transport time is too long in this field range to be resolved with this method.
However, it has been shown previously that even weakly coupled superlattices
exhibit a negative differential drift velocity in this regime.\cite{Grahn-PRB91}
In addition to the $C_1 \rightarrow C_j$ resonances due to $\Gamma$-subbands in
the GaAs wells and the exponential nonresonant background, a pronounced
step-like increase of the peak photocurrent is observed at about 10~V, which is
assigned to enhanced transport through the lowest $X$-level in the AlAs 
barriers. The lowest $X$-states
of the AlAs barriers are energetically located in such a way that transport
through these $X$ states becomes possible at voltages between the
$C_1 \rightarrow C_2$ and $C_1 \rightarrow C_3$ resonances. In the drift velocity
vs field characteristics, this $\Gamma \rightarrow X$ resonance leads to a
step-like increase of the drift velocity as indicated by the dotted line in
Fig.~\ref{slwfield}(b).

The voltages of the observed resonances can be compared to the calculated
energy levels listed in Table~\ref{samples}. For example, the bias voltages
$U_{bias}$ for $\Gamma \rightarrow \Gamma$ resonances are calculated using
\begin{equation}
U_{res}(C_1 \rightarrow C_j)=\frac{N~(C_j-C_1)}{e}\;,
\label{resonances1}
\end{equation}
where $C_1$ denotes the energy of the injecting conduction subband and
$C_j$ the energy of the conduction subband, in which the carriers are
tunneling into. $N$ refers to the number of periods and $e$ to the electron
charge. In the case of $\Gamma \rightarrow X$ resonances, however, the distance
between the levels is reduced to half the superlattice period. Therefore, an
additional factor of 2 has to be included, i.e.,
\begin{equation}
U_{res}(C_1 \rightarrow X_j)=2~\frac{N~(X_j-C_1)}{e}\;,
\label{resonances2}
\end{equation}
where $X_j$ denotes the energy of the receiving state in the barrier.
The voltages for the $\Gamma$ resonances $C_1 \rightarrow C_2$ and
$C_1 \rightarrow C_3$ are calculated using Eq.~(\ref{resonances1})
resulting in 5.4 and 14.6~V for the 9.0/4.0 sample, respectively.
Using Eq.~(\ref{resonances2}), we obtain a resonance voltage of 8.2~V
for the $C_1 \rightarrow X_1$ resonance in this sample. All values
agree well with the observed values. Note that all observed resonances
appear at slightly higher voltages due to field inhomogeneities and
screening effects.

To investigate the electric field inside the superlattice, photoluminescence
(PL) measurements have been performed.\cite{Grahn-jjap95,Grahn-PRB90} Due to
the quantum-confined Stark effect (QCSE), the PL from regions with strong
electric fields is red-shifted with respect to the PL from low-field regions.
The experiments reveal that in both plateau regions of this sample the PL
consists of two peaks with their respective intensity depending on the applied
bias voltage. This Stark-splitting of the PL signal is taken as direct evidence
for the existence of two distinct electric-field domains, which are formed
because of NDV and current conservation. For the first plateau the
consequence of this conservation law is depicted by the dashed line in
Fig.~\ref{slwfield}(b). Clearly, the electric field of the low-field domain
($F_{L}$) corresponds to the $C_1 \rightarrow C_1$ resonance peak, while the
field strength of the high field domain ($F_{R}$) lies on the next rising
branch of the $v(F)$ characteristics slightly below the
$C_1 \rightarrow C_2$ resonance.\cite{Grahn-PRL91,Kwok-PRB94}
Applied to the second plateau, the current conservation law (dashed lines in
Fig.~\ref{IV6750}) shows that the high field domain in this voltage region
forms as a consequence of resonant transport through the $X_1$ level in the
barriers ($C_1 \rightarrow X_1$ transport). In contrast to previous
work,\cite{Zhang-APL8-94} where this new formation mechanism was considered
for the first time, this measurements provide the first direct evidence relating
$X$-levels in the barriers to domain formation. However, it should be mentioned
that it is not very surprising to observe domain formation due to transport
through the $X_1$ level. For domain formation to occur, only a {\em minimum of
the $v(F)$ curve} is required in conjunction with a sufficiently high carrier
density.\cite{WAC96} In that sense the subject of domain formation and conditions
for oscillations can be discussed simply by looking at the $v(F)$
curve as discussed in a companion paper.\cite{WAC96}

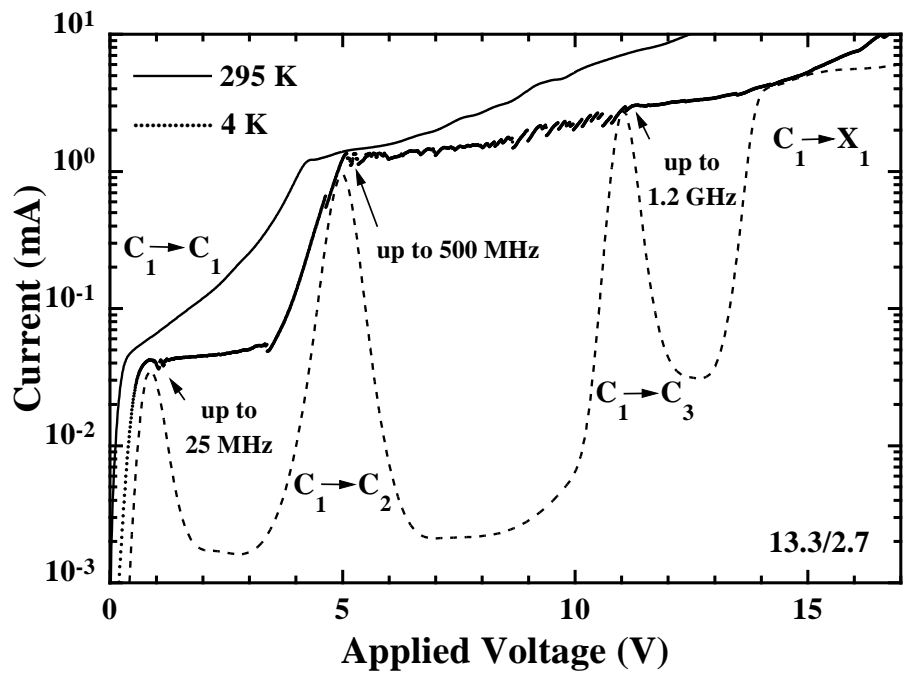
\begin{figure}
\caption{Current-voltage characteristics at 4 and 295 K for the 13.3/2.7
sample. The dashed line shows schematically the expected drift velocity vs
field characteristic of this sample for a homogeneous field.}
\label{IV6173}
\end{figure}

In the 13.3/2.7 sample the calculated $C_1 \rightarrow X_1$ resonance ($13.7~$V)
is located between the $C_1 \rightarrow C_3$ and $C_1 \rightarrow C_4$ resonances,
which are calculated to be at 9.5 and 17.9~V, respectively. Consequently, the
I-V characteristic exhibits three plateaus as can be seen in Fig.~\ref{IV6173}.
The thin dashed line has been added to indicate the positions of the transport
maxima for homogeneous fields. Unlike in the 9.0/4.0 sample, the PL in the plateau
regions shows no Stark splitting indicating that the field distribution inside the
SL is less sharply defined. A possible explanation of this behavior
is the existence of rather broad resonances in this weakly doped superlattice.
However, the I-V characteristics shows some clear signs of domain
formation (e.g., current branches\cite{Grahn-PRL91,Kastrup-APL94}), but
the differences between the electric field values of these domains seem
to be rather small. Since the QCSE is quadratic in the electric field,
the peaks of the two domains cannot be resolved in the PL experiments.
Because stable electric field domains have been investigated
intensively over the last years, we will not discuss the stationary regime
any further. We will rather focus on the non-stationary behavior instead.

\subsection{Oscillations in different plateaus}

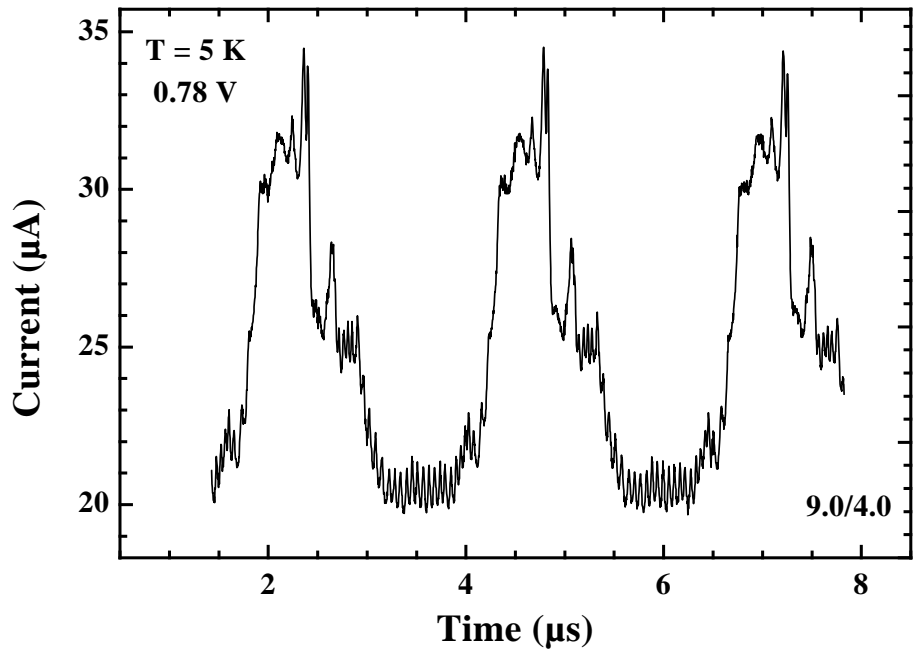
\begin{figure}
\caption{Current self-oscillations at 0.78 V in the first plateau of the
doped 9.0/4.0 sample at $4~$K.}
\label{trace6750}
\end{figure}

The absence of a regular pattern of current branches in the plateaus is a
first hint that the field distribution inside the SL may not always be
stable. Indeed, the current, e.g., in the lower plateau of the 9.0/4.0 sample
(Fig.~\ref{trace6750}), shows a strongly time-dependent component, which has
been attributed to an oscillatory instability of the field distribution
inside the superlattice.\cite{Kastrup-PRB95,Grahn-jjap95} During one cycle
of the oscillation a monopole forms near the cathode, drifts through the
superlattice towards the anode and dissolves, while a new monopole is
formed.\cite{Grahn-jjap95} This type of oscillation is denoted
the first oscillation mode.

It is interesting to note that the displayed trace also contains a number of
small current spikes. Supported by simulations, voltage turn-on measurements
have shown that these small spikes are due to well-to-well hopping of the
domain boundary moving through the superlattice.\cite{Kastrup-PRB96} Thus,
the number of current spikes indicates how many superlattice periods are
involved in the oscillation. For the displayed oscillation this leads to a
number close to thirty, which allows to speculate that for this particular
voltage the domain boundary crosses about three quarters of the superlattice
during one cycle of the oscillation.

To analyze the oscillations in more detail, we have measured the voltage
dependence of the oscillation frequency within the plateaus. The Fourier
transform\cite{Kastrup-PRB95} exhibits a series of frequency maxima, of which
we take the fundamental frequency as the oscillation frequency of the
field distribution.

\subsubsection{Voltage independent frequency}

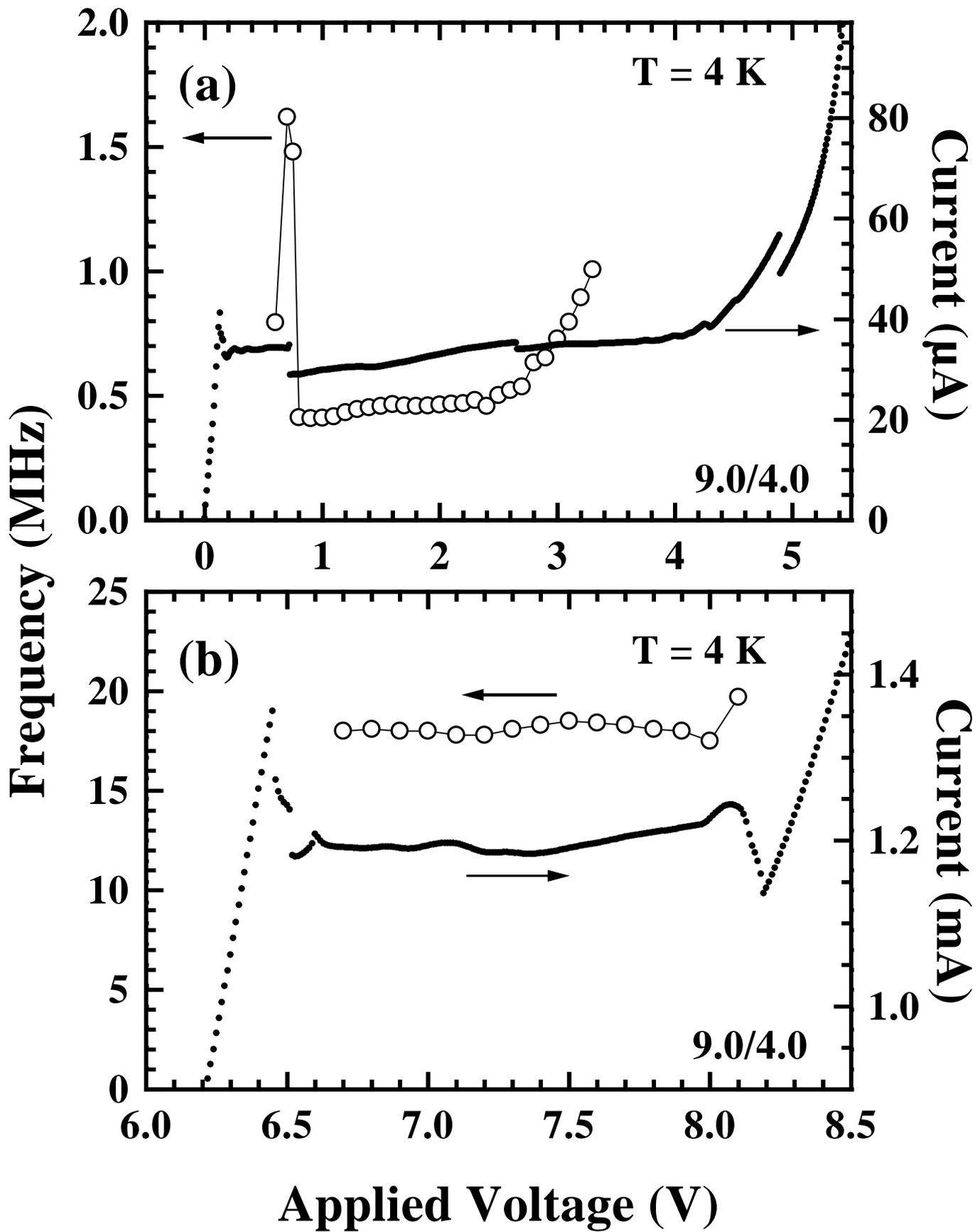
\begin{figure}
\caption{Voltage dependence of the oscillation frequency within the first
(a) and second (b) plateau of the doped 9.0/4.0 sample at $4~$K. The
frequencies are indicated
as open circles. For comparison the time-averaged I-V characteristics are
included
labeled by dots.}
\label{freq6750}
\end{figure}

The voltage dependence of the oscillation frequency in the two plateaus of
the 9.0/4.0 sample is plotted in Fig.~\ref{freq6750} showing that the observed
frequencies in the first plateau are weakly voltage dependent, while in the
second plateau there is no dependence on the applied voltage at all.
The noticeable increase of the frequency with voltage
at the end of the first plateau is due to the charge monopole having less
room to travel from the beginning of the superlattice to the position of
the high field domain. In both plateaus there are regions, where sinusoidal
oscillations could not be observed. In these regions the current contains
a time dependent component, which is not periodic and
therefore difficult to measure. In particular the discontinuities in the
frequency at the beginning of the first plateau suggest that the
oscillations may become chaotic, when the charge monopole oscillates near the
contacts. While chaotic behavior has been shown to exist theoretically in
the case of an external driving frequency,
\cite{Bulashenko-PRB95,Bulashenko-PRB96}
the boundary regions of the plateaus are still under investigation in this
respect. Very recently driven and undriven chaos have been observed in this
system.\cite{Zhang-PRL96} Note that chaotic behavior has also been observed
at the onset of trap-dominated Gunn oscillations in ultra-pure p-Ge.\cite{Kahn-PRB92}
This behavior is due to intermittent switching between small and large amplitude
current oscillations,\cite{Kahn-PRB92} although its theoretical interpretation
remains uncertain.\cite{Bergmann-PRB96}

\subsubsection{Voltage dependent frequency}

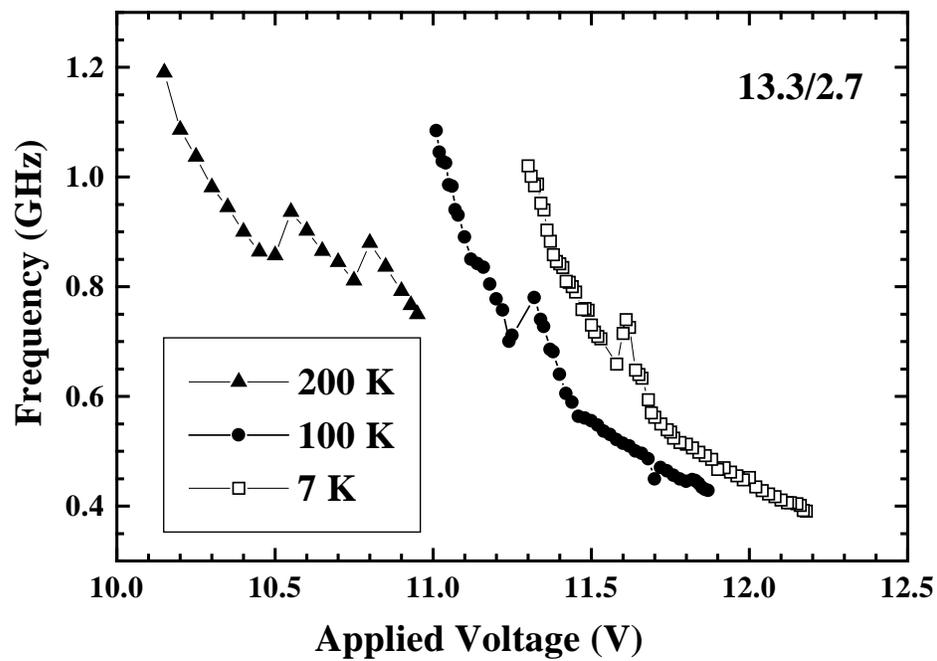
\begin{figure}
\caption{Oscillation frequency vs applied voltage in the third plateau
of the 13.3/2.7 sample for three different temperatures.}
\label{HF6173}
\end{figure}

In contrast to the 9.0/4.0 sample, the frequencies of the oscillations in the
13.3/2.7 sample exhibit a pronounced voltage dependence as shown in
Fig.~\ref{HF6173} for three different temperatures in the third plateau.
The oscillations begin to appear close to the resonance maximum and decrease
monotonically with increasing voltage. Furthermore, in this sample oscillations
are only detected near the beginning of the plateaus in the region of NDV.
This observation not only applies to the third plateau, but also to the
first and second plateaus. Thus, there is a clear difference between the
oscillations observed in these two samples. Unlike in the 9.0/4.0 sample, the
time-resolved PL does not show an oscillating
two-domain solution,\cite{Grahn-jjap95} but merely an intensity modulated
broad line. These two differences are taken as a signature for the existence
of a different oscillatory mode in this sample, which is not related to
well-defined electric field domains.

In Fig.~\ref{GHz465} the experimental results on the 9.0/1.5 sample
are shown. In this case we used a 25~$\mu$m mesa in order to keep the current
at a reasonable level. Due to the stronger coupling, GHz oscillations are
already observed in the lowest plateau. In addition to a strong
voltage dependence of the frequency, there is also an abrupt jump to lower
values. Higher plateaus could not be measured due to the large increase in
current at about 8~V. Time-resolved PL spectra on a 120~$\mu$m mesa do not
show two well-separated lines. However, intensity oscillations with the same
frequencies are clearly observed. Current oscillations on these large mesas
are also measured using pulsed electric fields to avoid the heating of the
sample.

\subsection{Temperature dependence}

Fig.~\ref{IV6173} shows the current voltage characteristic of the 13.3/2.7
sample at room temperature. From the fact that it still shows the
plateau-like features related to resonant tunneling one can speculate that
the oscillations may also be found at higher temperatures. In Fig.~%
\ref{HF6173} the oscillations in the third plateau are observed up to 200~K,
while in Fig.~\ref{RT6173} tunable oscillations indeed exist even at 295~K
in the case of the second plateau. These results are very encouraging,
because they prove that a device such as a frequency generator based on the
recently found tunable oscillatory mode can in principle be manufactured.
However, in order to obtain well-defined resonances even at room
temperature, one probably has to switch to another material system with
stronger confinement. For example, superlattices based on InGaAs/InAlAs
material system exhibit pronounced plateaus even at room
temperature.\cite{Kawamura87} Therefore, it is expected that with properly
chosen parameters oscillators based on InGaAs/InAlAs superlattices can be
operated at room temperature.

\section{Theoretical interpretation: Three oscillatory modes}

In order to explain why the frequency depends in one case on the applied
voltage, while it is voltage independent in the other case, the condition
for the two oscillatory modes has to be investigated in more detail.
Numerical calculations on the following discrete drift model have been
performed for the electric field $F_i$ and 3D electron density
$\tilde{n}_{i}$ of the $i^{th}$ SL period as well as for the total current
density $J$ using the following
equations,\cite{Kastrup-PRB95,Grahn-jjap95,Bonilla-PRB94,Bonilla-ICPF95}
\begin{eqnarray}
F_{i} - F_{i-1} = \frac{e\,d}{\epsilon}\,
(\tilde{n}_{i} - N_{D})\;,          \label{poisson}\\
\epsilon\,\frac{dF_{i}}{d\tilde{t}} +
e\,\tilde{v}(F_{i})\, \tilde{n}_{i} = J\;,  \label{ampere} \\
\frac{1}{N}\,\sum_{i=1}^{N} F_{i} = \frac{\Phi}{N\, d}\;.
\label{bias}
\end{eqnarray}
Here $i= 1,\dots, N$ is the period index, $N_D$ the 3D doping density,
$d$ the SL period, $\epsilon$ the average permittivity, and $\Phi$
the applied voltage between the two SL ends. Eqn.~(\ref{poisson})-(\ref{bias})
have to be supplemented using the appropriate initial conditions for the
field profile $F_i$ as well as with a boundary condition for $F_0$, which
we take as
\begin{equation}
F_{1}(t) - F_{0}(t) =
c\, \frac{e \, d \, N_{D}}{\epsilon}\;,
\label{bc}
\end{equation}
where $c>0$ denotes a dimensionless charge accumulation within the first SL
period due to the excess doping outside the SL ($n^+$-$n$-$n^+$ structure).
We expect that the narrower the barriers are, the larger $c>0$ is, since the
overlap between wavefunctions of adjacent quantum wells is larger for
narrower barriers. To find a precise relation between the structure of the
$n^+$ region before the SL and $c$, microscopic modeling of the full 
$n^+$-$n$-$n^+$ structure (characterized in Ref.\ \onlinecite{Kwok-PRB94}) 
is needed. This is outside the scope of the present work. However, we 
can estimate the values of $c$ by comparing experimental and calculated
frequency versus voltage bias curves. Thus comparing Figs.~\ref{freq6750}, 
\ref{HF6173} and \ref{GHz465} with Fig.~\ref{Fig5.4}(b), we obtain that $c$ 
should be very small ($c<0.001$) for samples 9.0/4.0 and 13.3/2.7, whereas 
$c$ should be from 10 to 100 times larger for sample 9.0/1.5 (see the
discussion below). 

The $\tilde{v}(F)$ function is a phenomenological electron velocity, which is
proportional to the tunneling probability. The main transport mechanism in
weakly coupled superlattices is sequential resonant tunneling (SRT). Thus,
$\tilde{v}(F)$ has peaks at the applied voltages or electric field strengths
corresponding to the alignment of the energy levels of adjacent quantum
wells as given in Eq.~(\ref{resonances1}), which are always separated by a
region of NDV. The experimentally observed voltage dependence of the frequency
in samples 9.0/4.0 and 13.3/2.7 (cf. Figs.~5 and 6) can be explained in
terms of the actual shape of the $\tilde{v}(F)$ curve, the doping density in the
wells and the charge accumulation within the first period.
The crucial features in the $\tilde{v}(F)$ curve are the separation between the
peaks as well as their width. The position of the peaks can be approximated
by Eq.~(\ref{resonances1}). The width of the peaks and the values of the
minima of $\tilde{v}(F)$ depend on the particular scattering mechanisms, which are
present in the samples. However, it is reasonable to assume that the peaks
of the $\tilde{v}(F)$ curve for the 13.3/2.7 sample are wider than those of the
9.0/4.0 sample, since the narrower the barriers, the wider the energy
intervals for resonant tunneling.\cite{Inarrea-95} Therefore, we will compare
simulations of the same model using two different $\tilde{v}(F)$
curves and different doping densities. We shall use curves $\tilde{v}(F)$
with only one peak, corresponding to a $C_1 \rightarrow C_i$ resonance,
with $i=1,2,\ldots$. Thus we are only dealing with the self-sustained 
oscillations occurring in the plateau after the chosen resonance. Adding more 
peaks does not appreciably change the self-sustained oscillations on each 
different plateau, although new features may appear for highly doped SLs. In 
fact, there are stationary nonuniform solution branches with electric field 
profiles presenting simultaneous coexistence of three or more domains if a 
$\tilde{v}(F)$ curve with several peaks is considered.\cite{Grahn-PRB90,murzin} 

To compare experimental results belonging to three different samples to our
data, it is convenient to render the Eqs.~(\ref{poisson})-(\ref{bc})
dimensionless. Suppose we want to analyze time-periodic current oscillations
at a certain plateau of the current-voltage characteristic. We adopt as the
units of electric field and velocity the corresponding values at the peak of
the velocity curve $\tilde{v}(F)$ prior to the plateau, $F_{M}$ and
$\tilde{v}_M$, respectively. Thus, we have to scale the magnitudes in the
figures by a different factor for each sample and plateau. The time scale
factors, $s_t$, for the  9.0/4.0 (second plateau, after the
$C_1 \rightarrow C_2$ resonance), 13.3/2.7 (second plateau, after the
$C_1 \rightarrow C_2$ resonance), and 9.0/1.5 (first plateau, after the
$C_1 \rightarrow C_1$ peak), samples are 3.57, 0.98, and 0.30~ns, respectively.
The electric field scale factors, $s_E$, are on the same order of magnitude,
$10^5$, $4.4\times 10^4$, and $2.4\times 10^4$~V/cm. We set\cite{WAC96}
\begin{eqnarray}
E_i = \frac{F_{i}}{F_{M}}\, , \quad\quad n_i =
\frac{ \tilde{n}_{i}}{N_{D}}\, ,
\quad\quad I =  \frac{\tilde{J}}{e N_{D}\tilde{v}_{M}}\,
, \nonumber\\
v=\frac{\tilde{v}}{\tilde{v}_{M}}\, ,\quad\quad
 t = \frac{ \nu \tilde{v}_{M}\,\tilde{t}}{d}\, ,
\quad\quad \phi = \frac{\Phi}{N\,F_{M}\,d}\;,
\label{tilden}
\end{eqnarray}
where the dimensionless parameter $\nu$ is defined as
\begin{equation}
\nu=\frac{e N_{D} d}{\epsilon F_{M}}\;.
\label{adimdop}
\end{equation}

A typical self-sustained current oscillation (far from the voltage
bias corresponding to the onset or the end of the instability) is
caused by the generation, motion and annihilation of
domain walls, which are charge monopoles. \cite{Bonilla-ICPF95}
We want to estimate  the dependence of the
oscillation frequency or period $T_p$ on the bias $\phi$.
There exist three different oscillation modes depending on
the value of the bias:
\begin{itemize}
\item {\em First mode:} Oscillation of the electric field profile around an
almost uniform stationary state (Fig. \ref{profiles}(a)).
It occurs when the bias is just above the onset of the instability.
It can be shown that the oscillation frequency decreases as the bias
increases by numerically solving the problem of linear stability of
the stationary electric field profile as discussed in a companion
paper.\cite{WAC96}
\item {\em Second mode:} Monopole recycling (Fig. \ref{profiles}(b)),
which typically occurs when
the bias is larger. In this mode of oscillation, the period is the sum
of the formation time of a monopole (the time it takes each new charge
accumulation injected at the first SL period to form a monopole)
plus the time elapsed until another charge accumulation is injected.
New monopoles are formed when the current is close to its maximum value
($I\approx 1$, or $\tilde{J}\approx eN_D\tilde{v}_M$ in dimensional units)
and the monopole formation time is small compared
to the period of the oscillation. The bias interval, over which the second
oscillatory mode exists, can be shown to increase with sample doping or
with the number of SL periods.
\item {\em Third mode:} Oscillation of the electric field profile about
a nonuniform stationary state containing two electric field domains
(Fig. \ref{profiles}(c)). This mode is sometimes present in weakly doped
samples, which oscillate for larger biases. It then occurs near the end
of the oscillatory branch. In this third mode of oscillation, a domain
wall remains pinned at a given location and the field values at the low
and high field domains oscillate in antiphase. The shape of the current
oscillation is almost sinusoidal and the maximum current is clearly
below 1 (where the accumulation layers were injected from the injecting
contact for the mode discussed before). A similar oscillation mode has been
found in simulations of imperfect superlattices.\cite{WAC95b}
\end{itemize}

Next we report  the results of numerical simulations of the
self-sustained oscillations of the current in the discrete model
for different values of the dimensionless parameters.
The simulations yield self-sustained current-oscillations when the
doping $\nu$ is such that the corresponding stationary field
profile is clearly inhomogeneous, e.g., as in Fig.~\ref{Fig5.3}.
The current on the middle branch of stationary solutions is
significantly larger than $v(\phi)$. In these cases the bias has to
be within a certain interval $(\phi_{\alpha},\phi_{\omega})$ ($\phi_{\alpha}
>1$ is on the NDV branch of $v(E)$, and $\phi_{\omega}$ may or may not be
larger than the minimum $E_m$) to generate self-sustained current
oscillations. Both $\phi_{\alpha}>1$ and $\phi_{\omega}$ are increasing
functions of $c$. In Fig.~\ref{Fig5.3}(a), the oscillatory
behavior begins at $\phi=\phi_{\alpha}\approx 1.100$ via a supercritical
Hopf bifurcation. The amplitude of the oscillation increases with
bias following a square-root law and the frequency is almost constant.
At $\phi=\phi_{\omega}\approx 1.617$ the branch of oscillatory
solutions disappears via a second Hopf bifurcation. For longer SLs
with $N=200$ and the same doping, Fig.~\ref{Fig5.3}(e), or for a
50-well SL with larger doping, \ref{Fig5.3}(c), the end of the
oscillatory branch is different: a limit cycle collides with
the unstable fixed point from the middle branch of the Z-shaped
current-voltage characteristics and disappears. This bifurcation
scenario results in a decrease of the frequency down to zero,
when the collision takes place, while the amplitude is unchanged to
lowest order. The bias interval where the oscillatory branch exists,
$(\phi_{\alpha},\phi_{\omega})$, shrinks as $\nu$ decreases, and for
$\nu<\nu_0$ ($\nu_0\approx 0.073$ for $N=50$, $c=10^{-4}$), there is
no oscillatory solution. Close to the Hopf bifurcations the first
and third modes of oscillation are found. We no longer see recycling
and motion of domain walls. Instead, in all these examples, there is
an interval of bistability between the self-sustained oscillations
and the stable stationary lower branch of the Z-shaped characteristic
when $\phi\in (\phi_{\beta},\phi_{\omega})$. The bistability may be
hard to observe experimentally because the basin of attraction of the
self-sustained current oscillations is very small and most initial
field profiles evolve to stationary situations.
A similar type of bistability can be found when (i) wiggles in the
static current-voltage characteristic $I^*(\phi)$
appear, and (ii) the upper branch of a wiggle is unstable against
oscillatory behavior while the lower branch is stable.
See Fig.~1 of Ref.~\onlinecite{WAC96}
for $\nu=0.3$. Fig.~\ref{Fig5.3} shows the variation in magnitude and
frequency of the current oscillations as a function of the bias for
different doping values and number of SL periods.

To clarify the nature of the oscillations let us describe the
{\em second mode} of oscillations, which occurs for most biases.
Fig.~\ref{Fig5.1}(a) shows the field profiles at different times of a given
period of the current oscillations for a 50-well SL.
We can identify a field profile consisting of two well-formed
field domains at time (1). We define the field of the domain to
be the electric field $E_i$ at the position where $n_i$ has a local
minimum, i.e., the variation in the field is minimal. The domain wall
between two domains is a charge monopole containing an excess of
electrons ($n_i>1$). Its evolution towards the receiving contact
(until it disappears) is captured
by snapshots at times (2) to (4) in Fig.~\ref{Fig5.1}(a).
Let $j(t)$ be the instantaneous position of the domain wall
defined as the SL period for which $n_i$ is a local maximum. Except
for the time intervals, in which two domain walls are present
simultaneously, the constant voltage condition implies that
\begin{equation}
E_l(t)=E_h(t)-\frac{N}{j(t)}\left(E_h(t)-\phi\right)\, ,
\label{Eqel}
\end{equation}
where $E_l$ and $E_h$ are the values of the field in the low and
high field domains, respectively. Within the low field domain
there is a tiny inhomogeneity (contact layer) close to the injecting
contact, which is due to the boundary condition.\cite{WAC96}
As long as the low field domain is in the region of positive
differential velocity ($E_l<1$), the field profile is stable and the
contact layer follows adiabatically the current. The position $j(t)$
of the domain wall moves to the right with a certain velocity
$v_{mon}/\nu$, where $v_{mon}$ is always significantly less than 1
(the peak velocity in dimensionless units). Eq.~(\ref{Eqel}) implies
that $E_l$, $E_h$, and $I$ must increase with time in order to
fulfill the fixed voltage condition. When $E_l$ surpasses
1, a small charge accumulation is injected at $i=1$, which moves
to the right and grows until it becomes a well-formed charge
monopole. See the snapshots at times (3)
and (4) in  Fig.~\ref{Fig5.1}(a). The process is
repeated in time after the old monopole has disappeared and the
new one is well established. Note that two different monopoles
are present simultaneously during a certain part of the oscillation
period. For a long SL the displacement current
is quite small during most of each oscillation period, and
we have $E_l\approx E^{(1)}(I)$ and $E_h\approx E^{(3)}(I)$
[$E^{(1)}(I) < E^{(2)}(I) < E^{(3)}(I)$ are the three solutions
of the equation $v(E) = I$ for $I<1$. Note that the solutions
$E^{(1)}$ and $E^{(2)}$ disappear for $I>1$]. See Fig.~\ref{Fig5.2}
for a 200-period SL.
Notice that now the domains are well-established at times for which
the values of the current are different from those depicted in
Fig.~\ref{Fig5.1}(a). Even for a smaller
SL as the one in Fig.~\ref{Fig5.1}, the displacement current is small,
when the current is near its maximum value for each period [see the
time intervals between (1) and (2) in  Fig.~\ref{Fig5.1}(b) and
between (2) and (3) in  Fig.~\ref{Fig5.2}(b)]. Notice that the field
value on the low field domain is near the peak of the $v(E)$ curve
during these time intervals. This will be useful in what follows.

\subsection{Dependence of the oscillation frequency with voltage}
It is important to understand the bias dependence of the frequency
which could be used to tune the frequency of an eventual device in
agreement with the experimental observations of Section II.
The frequency might increase or decrease with bias depending on the
sample parameters $N$ and $\nu$, as shown in Figs.~\ref{Fig5.3}(b,d,f).
Let us start at the time $t=t_0$ when $E_1 = 1 + c\nu$ [then $v(E_0)=1>
v(E_1)$, $E_l \approx 1$], the contact layer looses its stability, and
a charge accumulation is injected at $i=1$ ($t_0\approx 129$ in
Fig.~\ref{Fig5.2}). One period of duration $T_p$
is completed at $t=t_0+T_p$ when the next charge accumulation
is injected. At this instant, the position of the charge monopole
is given by Eq.~(\ref{Eqel}) with $E_l=1$ and $E_h\approx E^{(3)}(1)$
which yields
\begin{equation}
M:=j(t_0+T_p)=\frac{E^{(3)}(1)-\phi}{E^{(3)}(1)-1}\, N .\label{EqM}
\end{equation}
Thus, the charge accumulation has to travel a distance $M$
in the time $T_p$. During the monopole formation time $T_f$,
$E_l\approx 1$, and the mean velocity of the charge accumulation is
$\nu^{-1}$. On the other hand, the same charge accumulation moves at a
smaller mean velocity equal to $v_{mon}/\nu$ (with $v_{mon}<1$) for
the rest of the period, $T_p - T_f$, once it has become a charge
monopole. We therefore have
\begin{equation}
M=\frac{1}{\nu}T_f+\frac{v_{mon}}{\nu}\, (T_p-T_f) .
\end{equation}
This gives
\begin{equation}
T_p=\frac{\nu M}{v_{mon}}-T_f\left(\frac{1}{v_{mon}}-1\right) .
\label{Eqperiod}
\end{equation}
We now estimate the monopole formation time $T_f$.
The new charge accumulation $\delta n$ travels with a velocity
of the order of $v(E_l)/\nu \approx 1/\nu$ and is amplified in time
via
\begin{equation}
\frac{d\delta n(t)}{dt} =-v'(E_l)\,\delta n(t)\label{Eqamplification}
\end{equation}
with an initial value $\delta n(t_0)\sim c$. At time (3) of
Fig.~\ref{Fig5.2}, $t_0+T_f$, $\delta n$ is so large
that it separates two different well-formed field domains. Then
we may consider that a new charge monopole has been born.
This charge monopole travels at the
velocity $v_{mon}/\nu$ mentioned above and it sharpens as it moves; see
the point marked by (4) in the Fig.~\ref{Fig5.2}(a).
This stage lasts until the fields before and after the monopole
reach the values $E=E^{(1)}(I)$ and $E=E^{(3)}(I)$, respectively, and
we are back to situation (1) having completed one period.
Mathematically, this behavior can be well described by an asymptotic
analysis in the continuum limit ($\nu\to 0, N\nu=const$). \cite{BON95b}

In order to estimate $T_f$, we note that it is determined
by the condition $\delta n(t_0+T_f)/\delta n(t_0)=a$
where $a\sim 1/c$. Then Eq.~(\ref{Eqamplification}) yields
\begin{equation}
\log(a) = - \int_{t_0}^{t_0+T_f} v'(E_l(t))\, dt .
\end{equation}
Now we obtain $E_l(t)$ from Eq.~(\ref{Eqel}) with
$E_h\approx E^{(3)}(1),\, j(t)=M+v_{mon}(t-t_0)/\nu$. Up to the
first order in $(t-t_0)$, this yields
\begin{equation}
E_l(t)=1+\frac{N \, v_{mon} \, [E^{(3)}(1)-\phi]}{\nu M^2} \, (t-t_0) .
\end{equation}
Linearizing $v'(E_l)\approx -|v''(1)|(E_l-1)$, we obtain
\begin{equation}
T_f=\sqrt{\nu M}\cdot\sqrt{\frac{2 M\log(a)}{N \, v_{mon} \, |v''(1)|
\, [E^{(3)}(1)-1]}}\, .
\end{equation}
Now $M$ decreases as $\phi$ increases, which can be seen from Eq.~(\ref{EqM}).
As $v_{mon}<1$, Eq. (\ref{Eqperiod}) indicates that there are two competing
mechanisms influencing the dependence of $T_p$ with $\phi$. In general
the monopole formation time $T_f$ is negligible compared to $T_p(\phi)$ in
the following cases: (i) for long SLs ($M$ in Eq.~(\ref{Eqperiod}) is then
large), (ii) when the values of $c$ are large (i.e., $a$ small), (iii) for
large doping $\nu$. In these cases, $T_p(\phi)$ should be a decreasing
function. We find that the frequency increases with $\phi$ for large
values of the quantities $N,\, \nu,\,$ and $c$, while it decreases otherwise.
These behaviors are illustrated in Figs.~\ref{Fig5.3} and \ref{Fig5.4}.

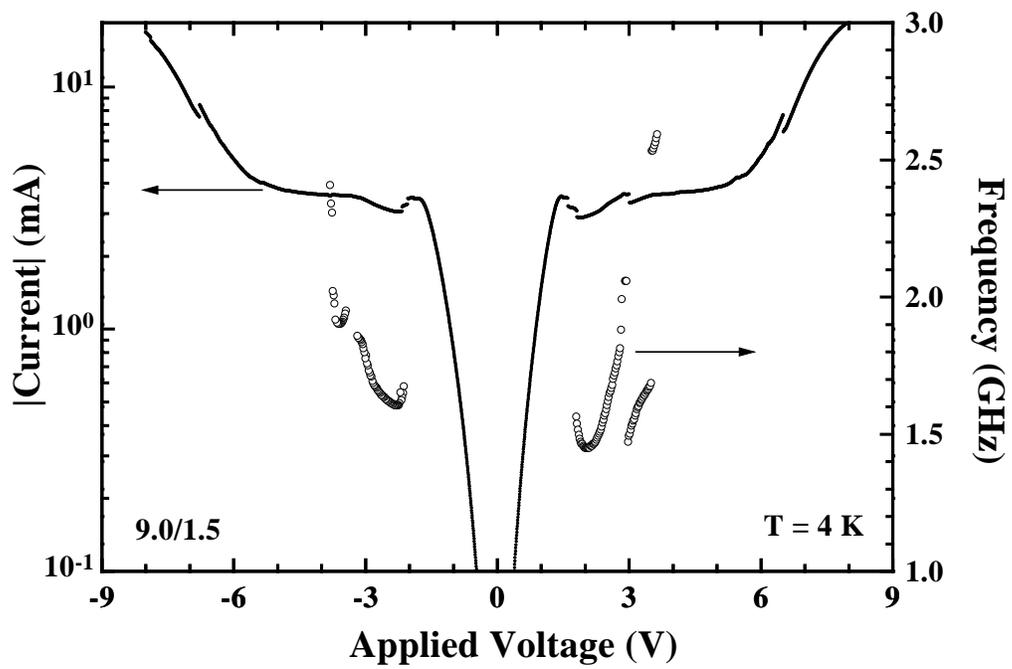
\begin{figure}
\caption{Voltage dependence of the oscillation frequency in the 9.0/1.5
sample at $4~$K. The frequencies are indicated as open circles. For comparison
the time-averaged I-V characteristic is included labeled by dots.}
\label{GHz465}
\end{figure}

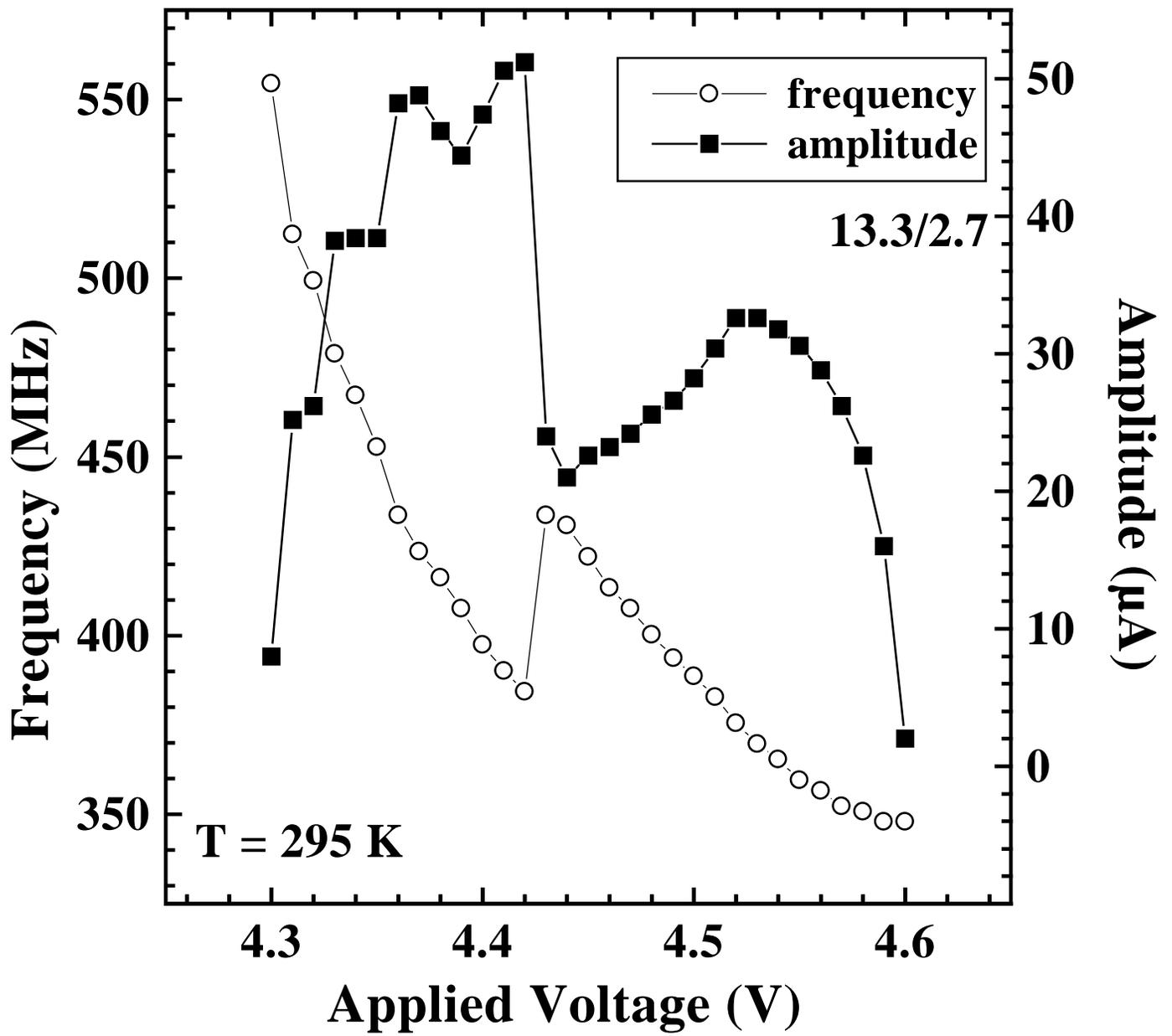
\begin{figure}
\caption{Oscillation frequency (open circles) and amplitude (squares) vs
applied voltage
in the 13.3/2.7 sample at room temperature in the second plateau.}
\label{RT6173}
\end{figure}

We can now interpret the experimental
observations according to our theoretical results.
\begin{itemize}
\item Sample 9.0/4.0 has narrow peaks, high doping with
respect to the others, and is short
(N=40). Therefore, the second oscillation mode dominates for most biases
with a short stage of monopole formation ($|v''(1)|$ is quite large).
During most of each oscillation period, the electric field on the low field
domain takes on values close to that of the resonant peak. This makes it
possible to distinguish two peaks in the PL data.
\cite{Kastrup-PRB95,Bonilla-PRB94,Kwok-PRB95} The
frequency is practically independent on the bias.\cite{Kastrup-PRB95}
This sample is modeled by the velocity curve shown in the inset of
Fig.~\ref{depen}(a). The results of the numerical simulation for current oscillations
and field profiles are also shown in
Fig.~\ref{depen}(a).

\item Sample 13.3/2.7 has wide peaks, low doping, and is also short
($N=50$). We would expect the first mode of oscillation to dominate,
which agrees with the experimental observation that the frequency is a
decreasing function of the bias. In this mode
the high field domain is never well established, and the low field domain
is off-resonance during most parts of each oscillation period
(cf. Fig.~\ref{depen}(b)). This explains
that no Stark-splitting of the PL signal is observed in the experiments for
this sample.

\item Sample 9.0/1.5 is similar to sample 13.3/2.7, except that its doping
is 2.5 times larger and the SL is shorter ($N=40$).
The velocity curve used in the simulations for this sample is shown in
the inset of Fig.~10(d). For this sample, Fig.~\ref{GHz465} seems
to indicate that the first two oscillation modes are important: for low
bias the first mode dominates, whereas for larger bias the second mode
does.  Fig.~\ref{GHz465} also shows that the frequency increases with
bias abruptly
after reaching its minimum value. Numerical calculations suggest that this
could correspond to having a larger excess charge in the first SL period
than in the case of sample 9.0/4.0. This could be expected for a SL with
narrower barriers (cf. Fig.~\ref{Fig5.4}).
\end{itemize}

\subsection{Self-sustained oscillations for SL with depletion contact layers}
Until now we have studied Eqs.~(\ref{poisson}) -
(\ref{bias}) with the boundary condition of Eq.~(\ref{bc}) and $c$ being
positive. Now we are going to describe what happens if $-1<c<0$, i.e.,
there are less electrons in the first well than the doping density
($n_1 -1 =c <0$; cf. Fig.~\ref{Fig6.2}).

As in the case of positive $c$, the oscillations are due to the generation,
motion and annihilation of domain walls connecting domains (which are
regions of almost uniform electric field approximately given by the zeros of
$I-v(E)$). The difference is that now the electric field profile is a monotone
decreasing function of the QW index: the high-field domain is close to the
beginning of the SL and the low-field domain extends to the end of the SL.
The domain walls are now charge-depletion layers, having less electrons
than the doping density. Let us describe one period of the current
oscillations for a long SL as the one in Fig.~\ref{Fig6.2} with $N=200$.
We will assume that initially (point marked with 1 in Fig.~\ref{Fig6.2}(a))
the field profile has two domains connected by a domain wall
($E=E^{(3)}(I_0)$ to the left of the domain wall and $E=E^{(1)}(I_0)$ to
the right of the domain wall with an initial value of the current
$I_{0}\in (v_m,1)$). The domain wall is approximately centered at $j/N=
Y=(\phi - E^{(1)}(I_0))/(E^{(3)}(I_0) - E^{(1)}(I_0))$ so that the
voltage bias condition of Eq.~(\ref{bias}) holds. The domain wall then
moves towards the end of the
superlattice with a speed close to the instantaneous value of the current.
The current decreases until a certain minimum value slightly smaller than
$v_m$ is reached. Then a new high-field domain is created (close to the
beginning of the SL) and the
current rises sharply as the two domain walls move towards the end of the SL.
When the current is near its maximum value, the old domain wall disappears as
the values of the field in the intermediate and rightmost domains coalesce.
Then one period of the oscillation is completed. All
these features can be understood by means of an asymptotic analysis to be
reported elsewhere. \cite{BON96} Some of
the basic ideas were sketched in Ref.~\onlinecite{Bonilla-ICPF95}.

\section{The superlattice oscillator as a tunable high-frequency generator}

Clearly, a superlattice oscillator based on the above described types
of oscillations may be exploited in a device, e.g., by operating the
superlattice as a frequency generator. In order to establish the use of
the newly found oscillations, it is important to know what maximum
frequency can be achieved. While experimental investigations in this
direction are still missing, it is clear that the ultimate frequency
limit will be connected with the maximum drift velocity, which can be
achieved without destroying the shape of the $v(F)$ curve.

To first order we assume that the oscillation frequency is determined
{\em only} by the transition time of the carriers through the sample.
This approach neglects a possible oscillation of the domain boundary
over a few periods only, but allows to give an estimate for the
oscillation frequencies from the current density $j$, the 3D doping density
$N_{D}$, and the length $L= N d$ of the sample using
\begin{equation}
f=\frac j{eL\,N_{D}}~.
\label{frequency}
\end{equation}
In Fig.~\ref{frequencies} the measured oscillation frequencies are plotted
versus the frequencies calculated using Eq.~(\ref{frequency}). For each
plateau in the 9.0/4.0 sample only one measured frequency is shown, while for
the 13.3/2.7 and 9.0/1.5 samples the measured minimum and maximum frequencies
are plotted. The figure shows a strong correlation between drift
velocity and oscillation frequency over several orders of magnitude. The
highest observed {\em miniband} drift velocities are on the order of
10$^7$~cm/s which, as Sibille et al.\cite{Sibille90} pointed out, for a
superlattice of 500$~$nm length translates into a fundamental frequency
of 200 GHz. However, the electric fields in the reported case cannot be
increased arbitrarily without breaking up the miniband, while for the subband
field-oscillator this limitation does not apply. Accordingly, the observed
frequencies become larger for higher subband resonances, which for
sufficiently thin barriers will push the drift velocities and the
frequencies to new limits. Looking at the oscillation frequency as a function
of barrier width, we note that a change of a factor of 2.67 in barrier width 
leads to a change in the oscillation frequency in the first plateau of
a factor of about 4000. Projecting this behavior for other material systems
to feasible barrier widths, we conclude that the frequencies could
reach the 100~GHz range.

Hence, in contrast to Gunn-effect devices, the oscillation frequency of the
superlattice oscillator is not only defined by the length of the sample,
but also by the parameters of the heterostructure. In this respect the
superlattice oscillator leaves much more room for optimization.
Additionally, a frequency generator based on a superlattice oscillator with
several plateaus may cover several orders of magnitude with one device. In
particular the tunability {\em within the plateaus} at high frequencies
allows to envision a tunable frequency generator based on resonant carrier
transport covering the whole GHz regime, which to our knowledge no other
device is capable of.

\section{Summary and Conclusions}

We have demonstrated that doped, weakly coupled semiconductor
superlattices exhibit current self-oscillations with GHz frequencies,
which can be tuned by simply changing the applied voltage.
The oscillations, which are observed in GaAs-AlAs superlattices,
are not limited to low temperatures, but in one case have been observed
at room temperature. They occur due to an unstable electric-field domain
formation. A theoretical modeling, both analytical and numerical,
demonstrates that three different oscillatory modes can occur, which
depend on the detailed shape of the drift velocity curve, the doping
density, the boundary condition, and the length of the superlattice.
In contrast to the Bloch-oscillator proposed by Esaki and
Tsu,\cite{Esaki70} the self-oscillations of the field distribution
are in most cases due to a formation, motion, and recycling of a
charge accumulation layer inside the superlattice.
In some cases the domain boundary is localized over a few quantum
wells, while the field strengths of the low and high field domain
oscillate. We would like to stress that the observed
oscillations are also different from the damped miniband oscillations
observed by Le Person et al.,\cite{LePerson92} because they occur only at
field strengths, where the miniband is already destroyed by Wannier-Stark
localization.

Considered from a macroscopic point of view in the sense that a traveling
charge accumulation layer generates the oscillation in the current, the
oscillations in doped superlattices are similar to the moving field domains
in Gunn diodes. In both cases the nonlinearity, which is responsible for
the instability, is generated by negative differential velocity. The
microscopic origin, however, is entirely different with important consequences
for the frequency limit and the tunability.

In conclusion, the subband superlattice oscillator has great potential for
applications as a tunable GHz oscillator. In order to achieve such a device,
it is necessary to fully understand the origin of the oscillations and
determine all possible oscillation modes. Furthermore, other material systems
such as InGaAs/InAlAs should be investigated in order to generate even higher
GHz oscillations at room temperature.

\section{Acknowledgments}

The authors would like to thank D. Bertram for the time-of-flight
experiments, R. Klann and S. Venakides for valuable discussions, A. Fischer
for sample growth, and E. Wiebecke and H. Kostial for sample preparation.
The work was supported in part by the DFG within the framework of Sfb 296, 
by the DGICYT grant PB94-0375, and by the EU Human Capital and Mobility 
Programme contract ERBCHRXCT930413. One of us (AW) acknowledges financial 
support from the Deutsche Forschungsgemeinschaft (DFG).

\begin{figure}
\caption{Current oscillations (left) and field profiles (right) for
$\nu = 0.1$, $c = 10^{-4}$, $N=50$ and different values of
$\phi=1.10$ (a), 1.25 (b) and 1.61 (c). Note that near the extrema
of the current oscillations, the field distribution
is close to the one of the stationary state.}
\label{profiles}
\end{figure}

\begin{figure}
\caption{(a) Stationary current-voltage characteristic (full line)
with maximum and minimum of the oscillating current (dotted lines)
$N=50$. The oscillatory branch begins at $\phi_{\alpha}\approx
1.100$ and ends at $\phi_{\omega}\approx 1.617$; the interval of bistability
begins at $\phi_{\beta}\approx 1.487$. (b) Fundamental frequency of the current
vs.\ the average electric field (bias) for $N=50$. (c,d) Same as (a,b)
but now $\nu=0.15$; $\phi_{\alpha}\approx 1.052$, $\phi_{\beta}\approx 1.556$,
$\phi_{\omega}\approx 1.797$. (e,f) Same as (a,b) but $N=200$;
$\phi_{\alpha}\approx 1.02$, $\phi_{\beta}\approx 1.622$,
$\phi_{\omega}\approx 1.868$. In all cases $c=10^{-4}$.}
\label{Fig5.3}
\end{figure}

\begin{figure}
\caption{(a) Electric field profiles at different times during one
period of the current oscillation depicted in the inset. (b) Time evolution
of the electric field values in the left, middle (when it exists) and right
domain of the SL. The corresponding values of $E^{(i)}(I(t))$, $i=1,2,3$,
are represented with dashed lines. Parameter values are
$\phi=1.25$, $c=10^{-4}$, $\nu=0.1$, and $N=50$.}
\label{Fig5.1}
\end{figure}

\begin{figure}
\caption{Same as in Figure 11 for $N=200$.}
\label{Fig5.2}
\end{figure}

\begin{figure}
\caption{(a) Frequency vs.\ bias diagrams for different values of the dimensionless
doping $\nu$ and $c=0.001$. (b) Frequency vs.\ bias diagrams for different values of
$c$ keeping $\nu=0.1$ and $N=50$ fixed.}
\label{Fig5.4}
\end{figure}

\begin{figure}
\caption{Current oscillations (left) and field profiles (right)
simulating samples 9.0/4.0 (a) and 13.3/2.7 (b).}
\label{depen}
\end{figure}

\begin{figure}
\caption{(a) Current density versus time when $c=-0.01$.
Electron density (b) and
electric field (c) profiles during one period of the current oscillation.
The numbers in (b) and (c) correspond to the times marked in (a). Parameter
values are $\phi=1.25$, $\nu=0.35$, and $N=200$. }
\label{Fig6.2}
\end{figure}

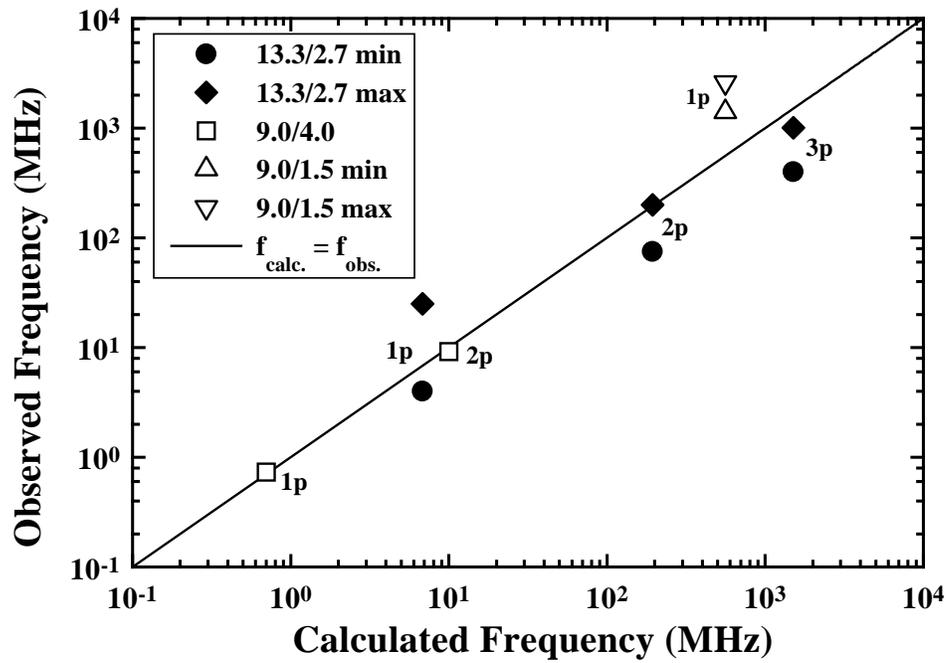
\begin{figure}
\caption{Calculated vs measured frequencies for the different plateaus in
the doped samples.
1p, 2p, and 3p denote the first, second, and third plateau, respectively.
For the 13.3/2.7 and 9.0/1.5 samples the minimum and maximum frequencies
for each plateau are
shown.}
\label{frequencies}
\end{figure}

\begin{table}
\caption{Parameters of the four investigated samples. $N$ denotes the
number
of periods and $N_{2D}$ the two-dimensional carrier density introduced by
doping.
$C_{i}$ and $X_{1}$ indicate the energy positions of the lowest
$\Gamma$-subbands in
the wells and the lowest $X$-subbands in the barriers, respectively. The
width
of the lowest conduction miniband is given by $\Delta_1$. The
energies of the subbands measured from the bottom of the
GaAs conduction band have been obtained by conventional envelope
function calculations.\label{samples}}
\begin{tabular}{c c c c c c c c c}
$d_{GaAs}/d_{AlAs}$ & $N$ & type & $N_{2D}$ & $C_1$ & $C_2$ & $C_3$ & $X_1$
 & $\Delta_1$
\\
(nm/nm) &  &  & (cm$^{-2}$) & (meV) & (meV) & (meV) & (meV) & (meV) \\
\tableline
13.3/2.7 & 50 & $n^{+}$-$n$-$n^{+}$ & $1\times 10^{10}$ & 23 & 94 & 212
& 159 & 0.13 \\
9.0/4.0 & 40 & $n^{+}$-$n$-$n^{+}$ & $1.5\times 10^{11}$ & 44 & 180 & 410
& 146 & $<0.1$ \\
9.0/4.0 & 40 & $p^{+}$-$i$-$n^{+}$ & - & 44 & 178 & 410 & 146 & $<0.1$ \\
9.0/1.5 & 40 & $n^{+}$-$n$-$n^{+}$ & $2.5\times 10^{10}$ & 44 & 180 & 411
& 199 & 3.7 \\
\end{tabular}
\end{table}

\begin{references}

\bibitem[a)]{JK}
present address: CompuNet GmbH, H\"orselbergstr. 7
D-81677 M\"unchen, Germany.

\bibitem[b)]{HTG}
permanent address: Paul-Drude-Institut f\"ur
Festk\"orperelektronik, Hausvogteiplatz 5-7,
D-10117 Berlin, Germany.

\bibitem[c)]{AW}
present address: Mikroelektronik Centret,
Technical University of Denmark, DK-2800 Lyngby, Denmark.

\bibitem{Ridley61}
B.~K.~Ridley and T.~B.~Watkins,
Proc. Phys. Soc. {\bf 78}, 293 (1961).

\bibitem{Gunn63}
J.~B.~Gunn,
Solid State Commun. {\bf 1}, 88 (1963).

\bibitem{Shur87}
M.~Shur, {\em GaAs devices and circuits}
(Plenum Press, New York, 1987), p. 173.

\bibitem{Higuera92}
F.~J.~Higuera and L.~L.~Bonilla,
Physica D {\bf 57}, 164 (1992).

\bibitem{Ruttan75}
T.~G.~Ruttan,
Electron. Lett. {\bf 11}, 294 (1975).

\bibitem{Bloch28}
F.~Bloch,
Z. Phys. {\bf 52}, 555 (1928).

\bibitem{Zener34}
C.~Zener,
Proc. R. Soc. London Ser. A {\bf 145}, 523 (1934).

\bibitem{Esaki70}
L.~Esaki and R.~Tsu,
IBM J. Res. Develop. {\bf 14}, 61 (1970).

\bibitem{Feldmann-PRB92}
J.~Feldmann et al.,
Phys. Rev. B {\bf 46}, 7252 (1992).

\bibitem{LePerson92}
H.~Le~Person, C.~Minot, L.~Boni, J.~F.~Palmier, and F.~Mollot,
Appl. Phys. Lett. {\bf 60}, 2397 (1992).

\bibitem{Hadjazi93}
M.~Hadjazi, J.~F.~Palmier, A.~Sibille, H.~Wang, E.~Paris, and F.~Mollot,
Electron. Lett. {\bf 27}, 648 (1993).

\bibitem{Kastrup-PRB95}
J.~Kastrup et al.,
Phys. Rev. B {\bf 52}, 13761 (1995).

\bibitem{Grahn-jjap95}
H.~T.~Grahn et al.,
Jpn. J. Appl. Phys. {\bf 34}, 4526 (1995).

\bibitem{Kastrup-SSE95}
J.~Kastrup, H.~T.~Grahn, K.~Ploog, and R.~Merlin,
Solid-State Electron. {\bf 40}, 157 (1996).

\bibitem{WAC96}
A. Wacker, M. Moscoso,  M. Kindelan, and L. L. Bonilla,
Phys. Rev. B {\bf 55}, 2466 (1997).

\bibitem{Schneider89}
H.~Schneider, K.~von Klitzing, and K.~Ploog,
Europhys. Lett. {\bf 8}, 75 (1989).

\bibitem{Grahn-PRB91}
H.~T.~Grahn, K.~von Klitzing, K. Ploog, and G.H. D\"ohler,
Phys. Rev. B {\bf 43}, 12094 (1991).

\bibitem{Grahn-PRB90}
H.~T.~Grahn, H.~Schneider, and K.~von Klitzing,
Phys. Rev. B {\bf 41}, 2890 (1990).

\bibitem{Grahn-PRL91}
H.~T.~Grahn, R.~J.~Haug, W.~M{\"u}ller, and K.~Ploog,
Phys. Rev. Lett. {\bf 67}, 1618 (1991).

\bibitem{Kwok-PRB94}
S.~H.~Kwok, R.~Merlin, H.~T.~Grahn, and K.~Ploog,
Phys. Rev. B {\bf 50}, 2007 (1994).

\bibitem{Zhang-APL8-94}
Y.~Zhang, X.~Yang, W.~Liu, P.~Zhang, and D.~Jiang,
Appl. Phys. Lett. {\bf 65}, 1148 (1994).

\bibitem{Kastrup-APL94}
J.~Kastrup et al.,
Appl. Phys. Lett. {\bf 65}, 1808 (1994).

\bibitem{Kastrup-PRB96}
J.~Kastrup, F.~Prengel, H.T.~Grahn, K.~Ploog, and E.~Sch{\"o}ll,
Phys. Rev. B {\bf 53}, 1502 (1996).

\bibitem{Bulashenko-PRB95}
O.~M.~Bulashenko and L.~L.~Bonilla,
Phys. Rev. B {\bf 52}, 7849 (1995).

\bibitem{Bulashenko-PRB96}
O.~M.~Bulashenko, M.~J.~Garc\'\i a, and L.~L.~Bonilla,
Phys. Rev. B {\bf 53}, 10008 (1996).

\bibitem{Zhang-PRL96}
Y.~Zhang, J.~Kastrup, R.~Klann, K.~H.~Ploog, and H.~T.~Grahn,
Phys. Rev. Lett. {\bf 77}, 3001 (1996).

\bibitem{Kahn-PRB92}
A.~M.~Kahn, D.~J.~Mar, and R.~M.~Westervelt,
Phys. Rev. B {\bf 46}, 7469 (1992).

\bibitem{Bergmann-PRB96}
M.~J.~Bergmann, S.~W.~Teitsworth, L.~L.~Bonilla, and I.~R.~Cantalapiedra,
Phys. Rev. B {\bf 53}, 1327 (1996).

\bibitem{Kawamura87}
Y.~Kawamura, K.~Wakita, and K.~Oe,
Jpn. J. Appl. Phys. {\bf 26}, L1603 (1987).

\bibitem{Bonilla-PRB94}
L.~L.~Bonilla, J.~Gal\'an, J.~A.~Cuesta,
F.~C.~Mart\'\i nez, and J.~M.~Molera,
Phys. Rev. B {\bf 50}, 8644 (1994).

\bibitem{Bonilla-ICPF95}
L.~L.~Bonilla, in
{\em Nonlinear Dynamics and Pattern Formation in Semiconductors
and Devices}, edited by F.-J.\ Niedernostheide (Springer
Verlag, Berlin, 1995), Chap.~1.

\bibitem{Inarrea-95}
J.~I\~narrea and G.~Platero,
Phys. Rev. B {\bf 51}, 5244 (1995).

\bibitem{murzin} 
Yu. A. Mityagin, V. N. Murzin, Yu. A. Efimov and G. K. Rasulova, 
Superlattices Microstructures {\bf 20}, in press (1996).

\bibitem{WAC95b}
A. Wacker, G. Schwarz, F. Prengel, E. Sch{\"o}ll, J. Kastrup, and
H.T. Grahn, Phys.~Rev.~B {\bf 52}, 13788 (1995).

\bibitem{BON95b}
L. L. Bonilla, M. Kindelan, M. Moscoso, and S. Venakides, 
SIAM J. Appl. Math. {\bf 57}(6), in press, cond-mat/9611239. 

\bibitem{Kwok-PRB95}
S. H. Kwok et al.,
Phys. Rev. B {\bf 51}, 10171 (1995).

\bibitem{BON96}
L. L. Bonilla, M. Kindelan, M. Moscoso, and A. Wacker, (unpublished).

\bibitem{Sibille90}
A.~Sibille, J.~F.~Palmier, H.~Wang, and F.~Mollot,
Phys. Rev. Lett. {\bf 64}, 52 (1990).

\end{references}
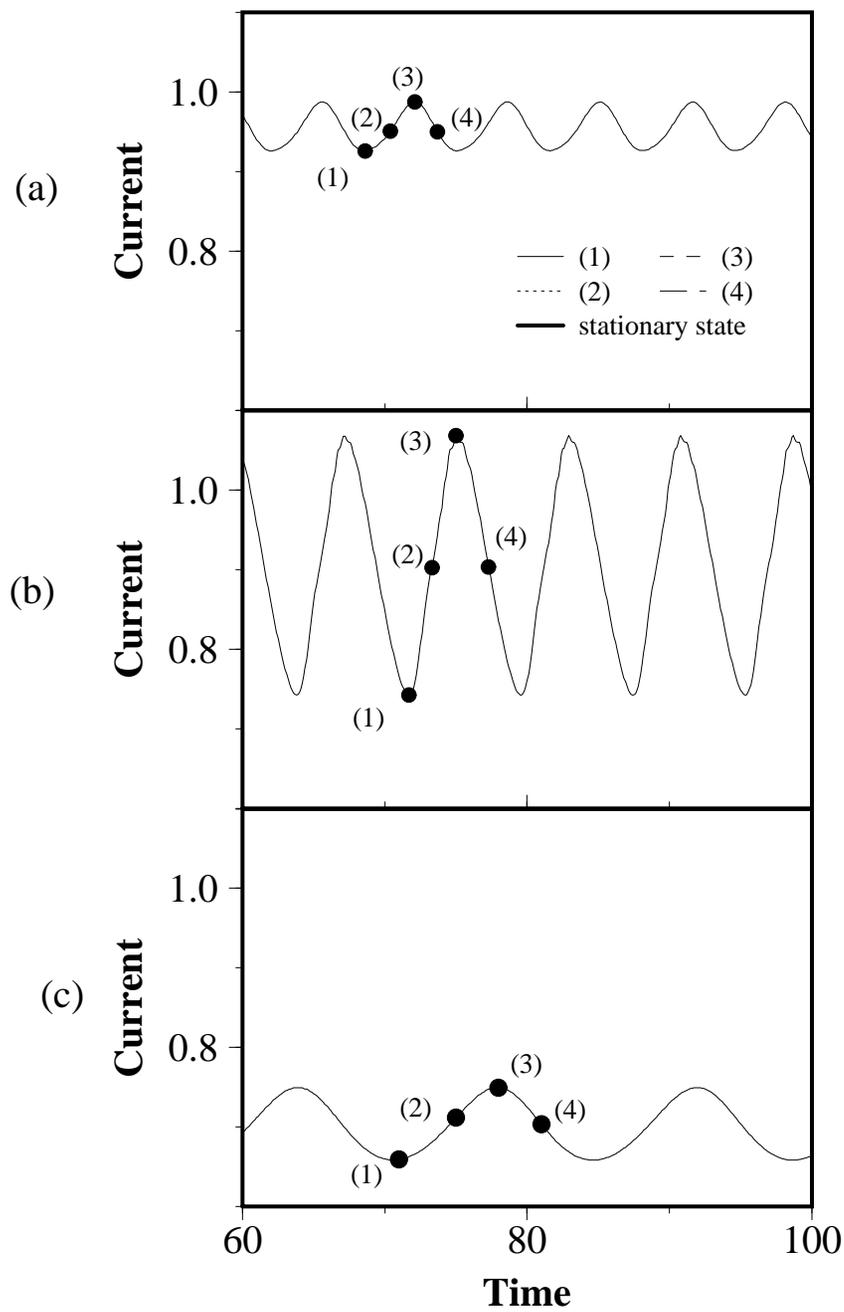
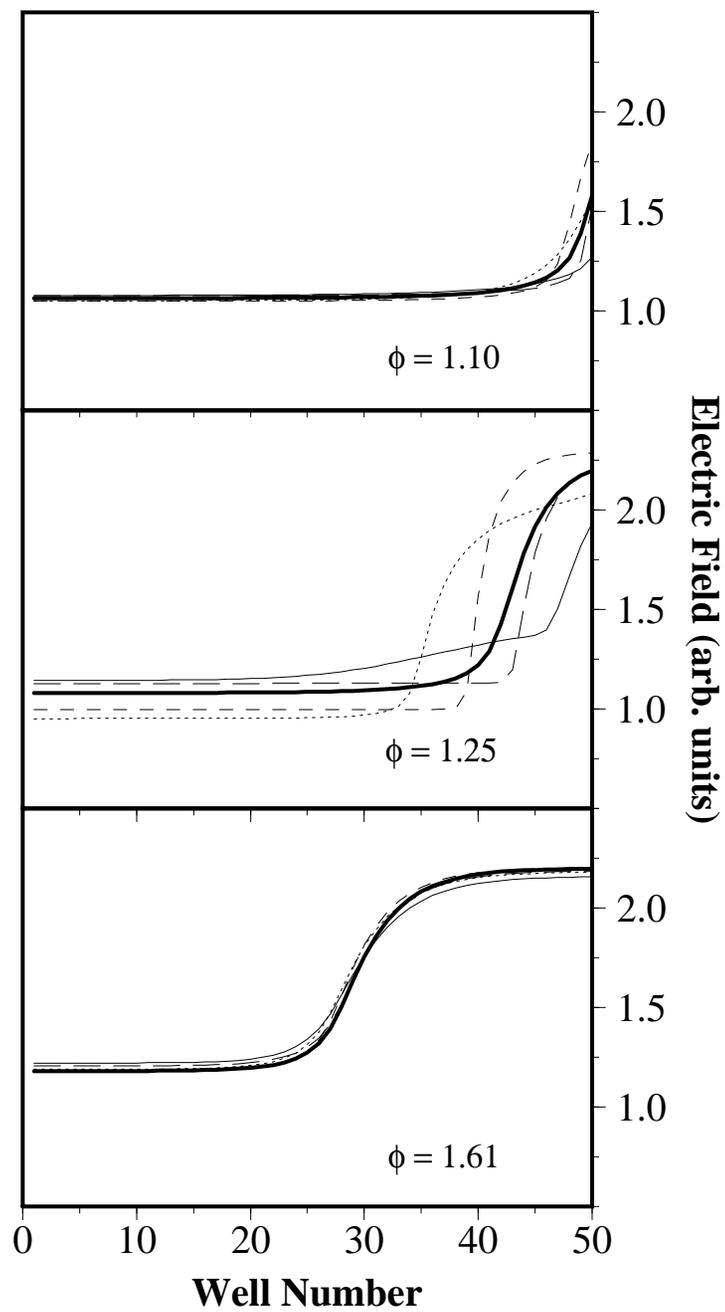
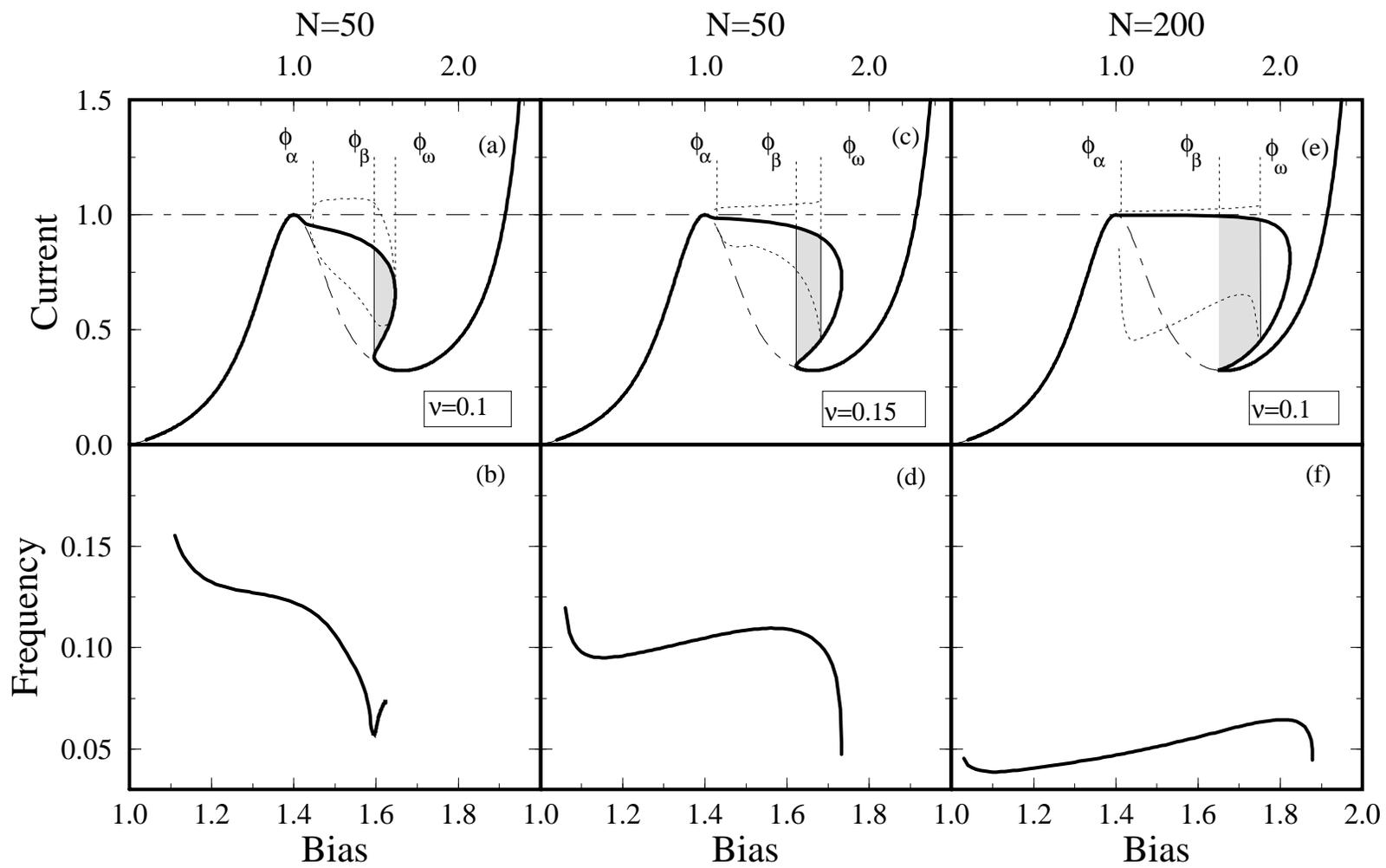
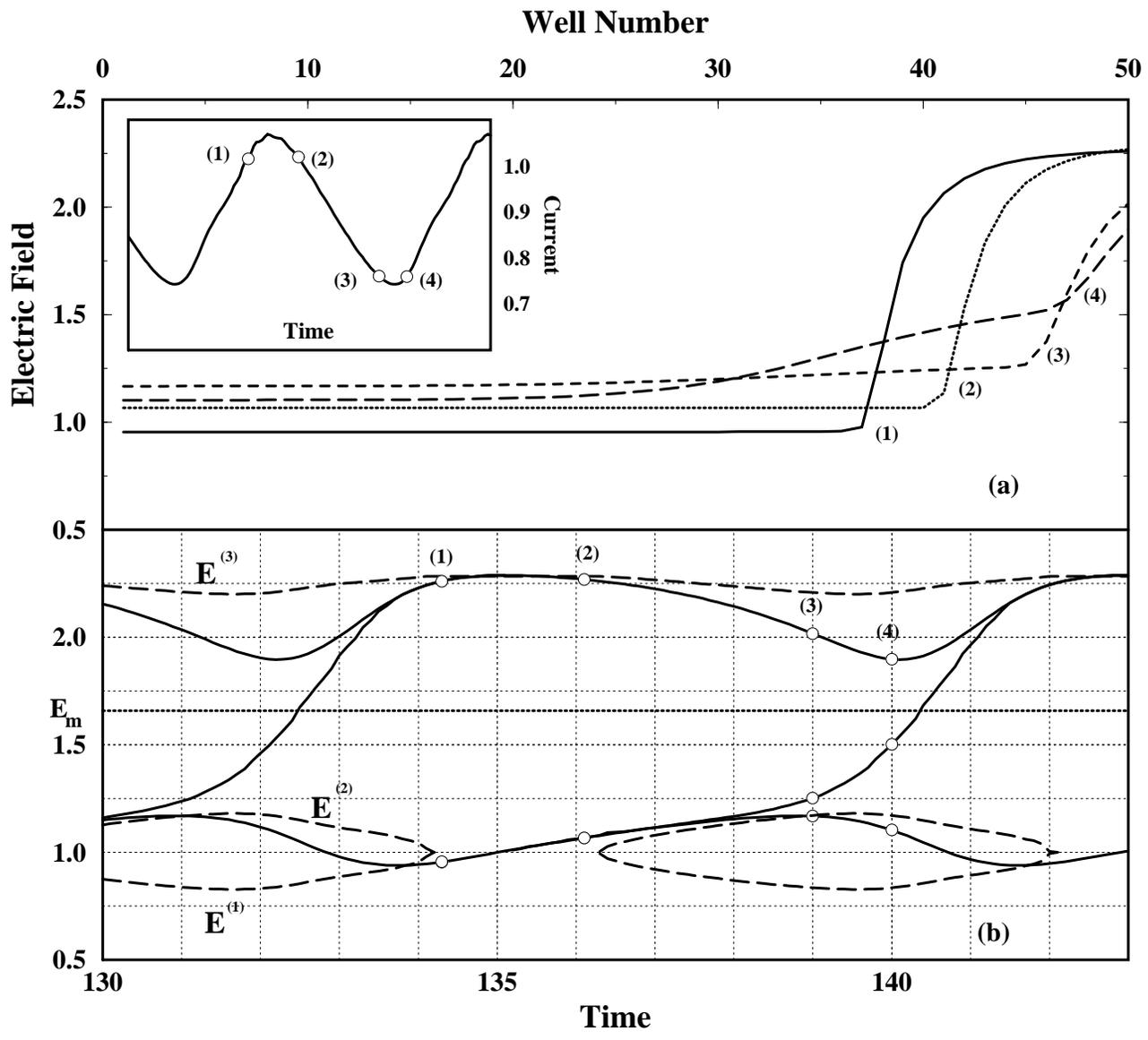
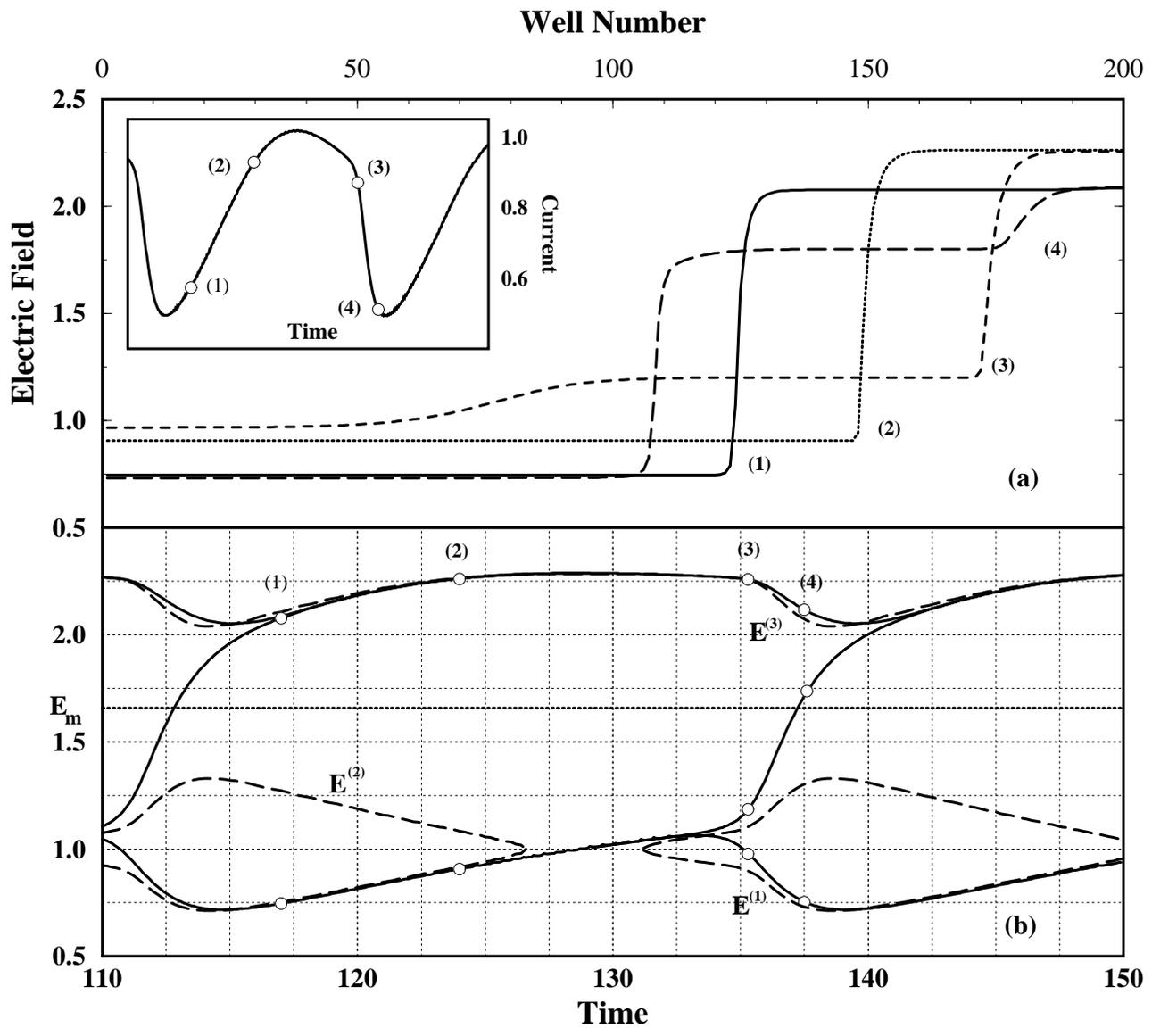
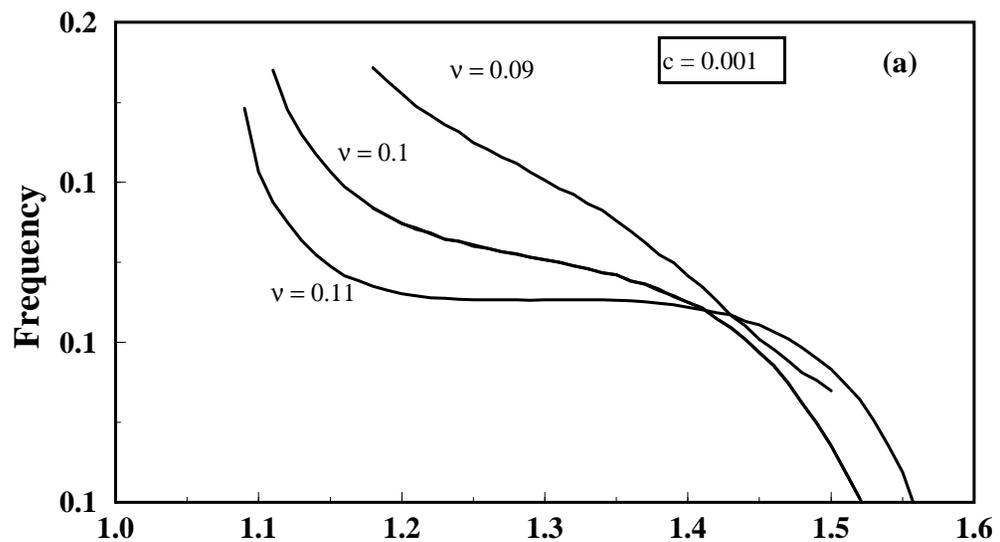
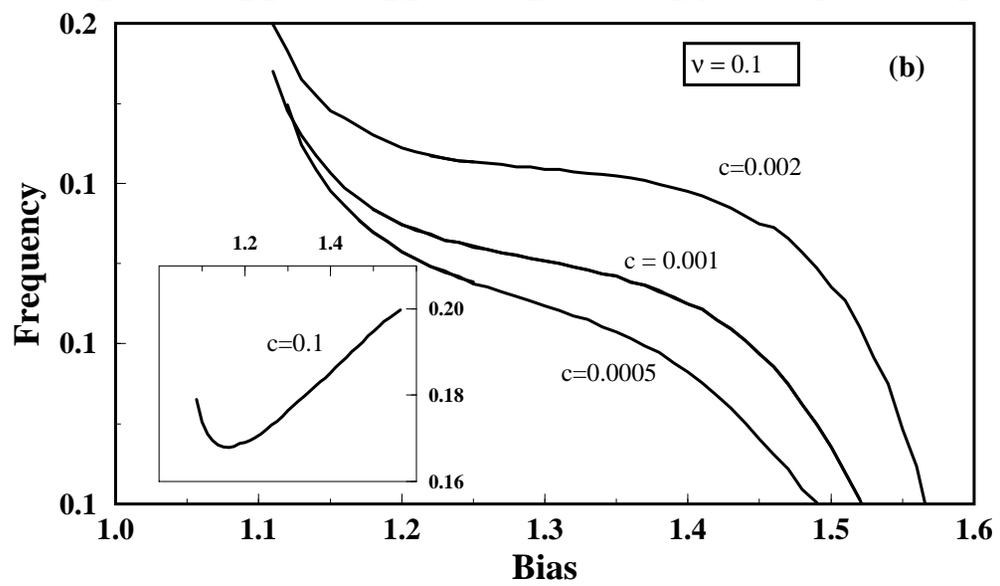
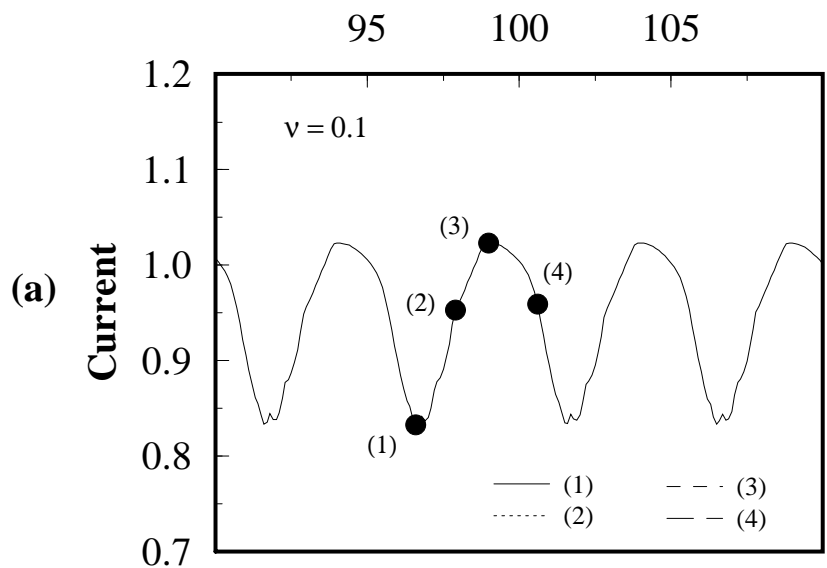
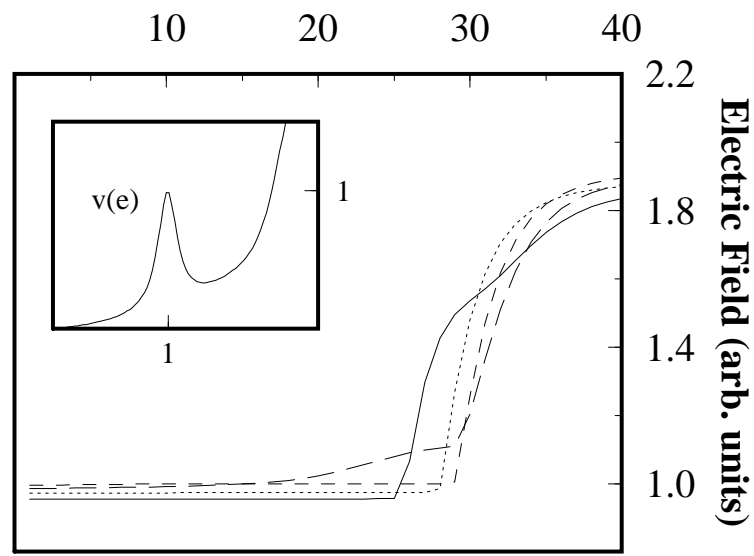
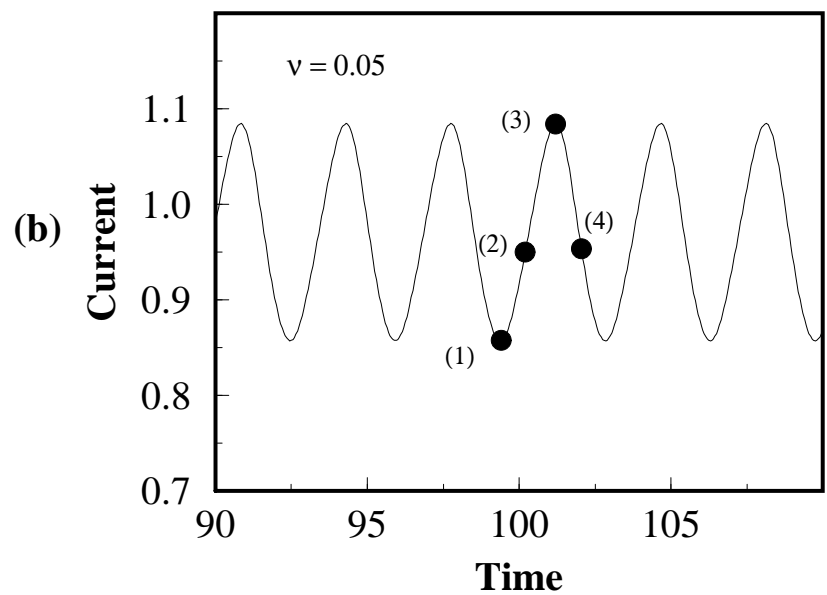
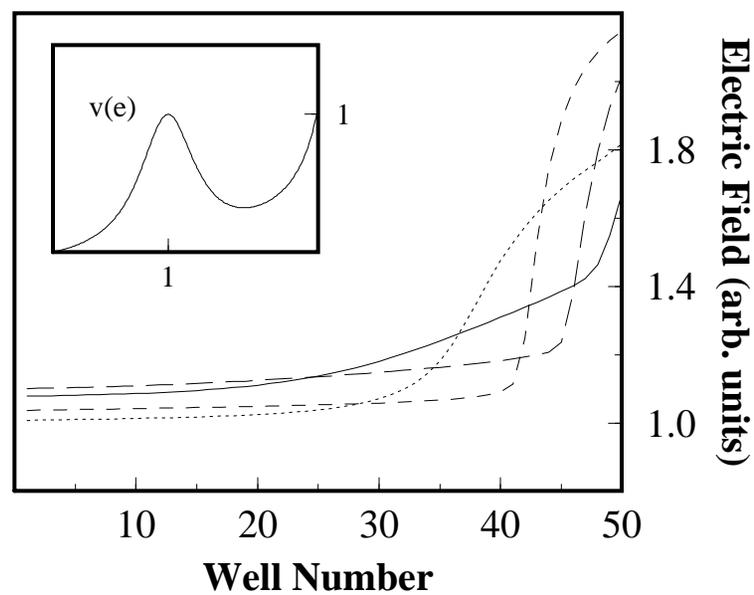
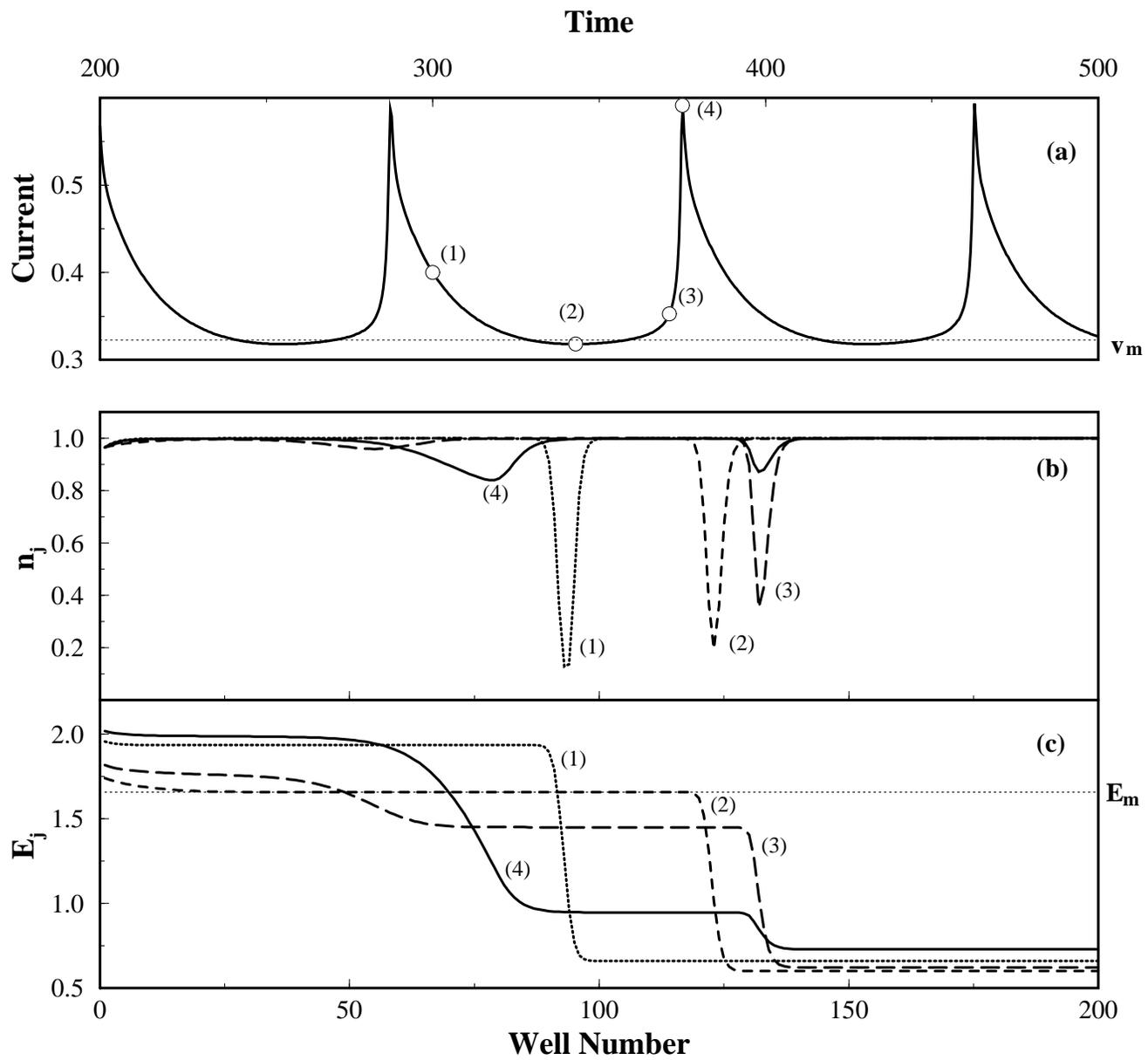
\end{document}